\documentclass[sigconf]{acmart}

\AtBeginDocument{%
  }

\setcopyright{acmlicensed}
\copyrightyear{2018}
\acmYear{2018}
\acmDOI{XXXXXXX.XXXXXXX}

\acmConference[Conference acronym 'XX]{Make sure to enter the correct
  conference title from your rights confirmation email}{June 03--05,
  2018}{Woodstock, NY}

\acmISBN{978-1-4503-XXXX-X/2018/06}

\usepackage{multirow}
\usepackage{framed}

\begin{document}

\title[\sysname{}: Capturing Workflows through Decision Structuring, Questioning, and Rationale Inference]{\sysname{}: Capturing Creative Workflows through Decision Structuring, In-Situ Questioning, and Rationale Inference}

\author{Kihoon Son}
\email{kihoon.son@kaist.ac.kr}
\affiliation{%
  \institution{School of Computing, KAIST}
  \city{Daejeon}
  \country{Republic of Korea}
}
\author{DaEun Choi}
\email{daeun.choi@kaist.ac.kr}
\affiliation{%
  \institution{School of Computing, KAIST}
  \city{Daejeon}
  \country{Republic of Korea}
}
\author{Tae Soo Kim}
\email{taesoo.kim@kaist.ac.kr}
\affiliation{%
  \institution{School of Computing, KAIST}
  \city{Daejeon}
  \country{Republic of Korea}
}
\author{Young-Ho Kim}
\email{yghokim@younghokim.net}
\affiliation{%
  \institution{NAVER AI Lab}
  \city{Seongnam}
  \country{Republic of Korea}
}
\author{Sangdoo Yun}
\email{sangdoo.yun@navercorp.com}
\affiliation{%
  \institution{NAVER AI Lab}
  \city{Seongnam}
  \country{Republic of Korea}
}
\author{Juho Kim}
\email{juhokim@kaist.ac.kr}
\affiliation{%
  \institution{School of Computing, KAIST}
  \city{Daejeon}
  \country{Republic of Korea}
}
\email{juho@skillbench.com}
\affiliation{%
  \institution{SkillBench}
  \city{Santa Barbara}
  \country{USA}
}
% \email{juho@skillbench.com}
% \affiliation{%
%   \institution{SkillBench}
%   \city{Santa Barbara}
%   \country{CA, USA}
% }

\renewcommand{\shortauthors}{Son et al.}
\newcommand{\sysname}[0]{\textsc{ClearFairy}}
\newcommand{\approachname}[0]{\textsc{Clear}}
\newcommand{\username}[0]{Lillie}
\newcommand{\revise}[1]{%
  \textcolor{red}{\fcolorbox{red}{red}{\textcolor{white}{R}}~#1}%
}

\definecolor{blockcolor}{HTML}{555555}
\definecolor{blockrule}{gray}{0.6}
\newenvironment{block}%
  {\list{}{\leftmargin=0.2in\rightmargin=0.2in}\item[]\color{blockcolor}}%
  {\endlist}
  
\newenvironment{prompt}{
  \vspace{5pt}
  \begin{framed}
    \scriptsize\ttfamily\color{blockcolor}
}{
  \end{framed}
}

\begin{abstract}
Capturing professionals’ decision-making in creative workflows (e.g., UI/UX) is essential for reflection, collaboration, and knowledge sharing, yet existing methods often leave rationales incomplete and implicit decisions hidden. To address this, we present the \approachname{} approach, which structures reasoning into cognitive decision steps—linked units of actions, artifacts, and explanations making decisions traceable with generative AI. Building on \approachname{}, we introduce \sysname{}, a think-aloud AI assistant for UI design that detects weak explanations, asks lightweight clarifying questions, and infers missing rationales. In a study with twelve professionals, 85\% of \sysname{}’s inferred rationales were accepted (as-is or with revisions). Notably, the system increased ``strong explanations''—rationales providing sufficient causal reasoning—from 14\% to 83\% without adding cognitive demand. Furthermore, exploratory applications demonstrate that captured steps can enhance generative AI agents in Figma, yielding predictions better aligned with professionals and producing coherent outcomes. We release a dataset of 417 decision steps to support future research.
\end{abstract}




\begin{teaserfigure}
   \centering
  \includegraphics[width=\textwidth]{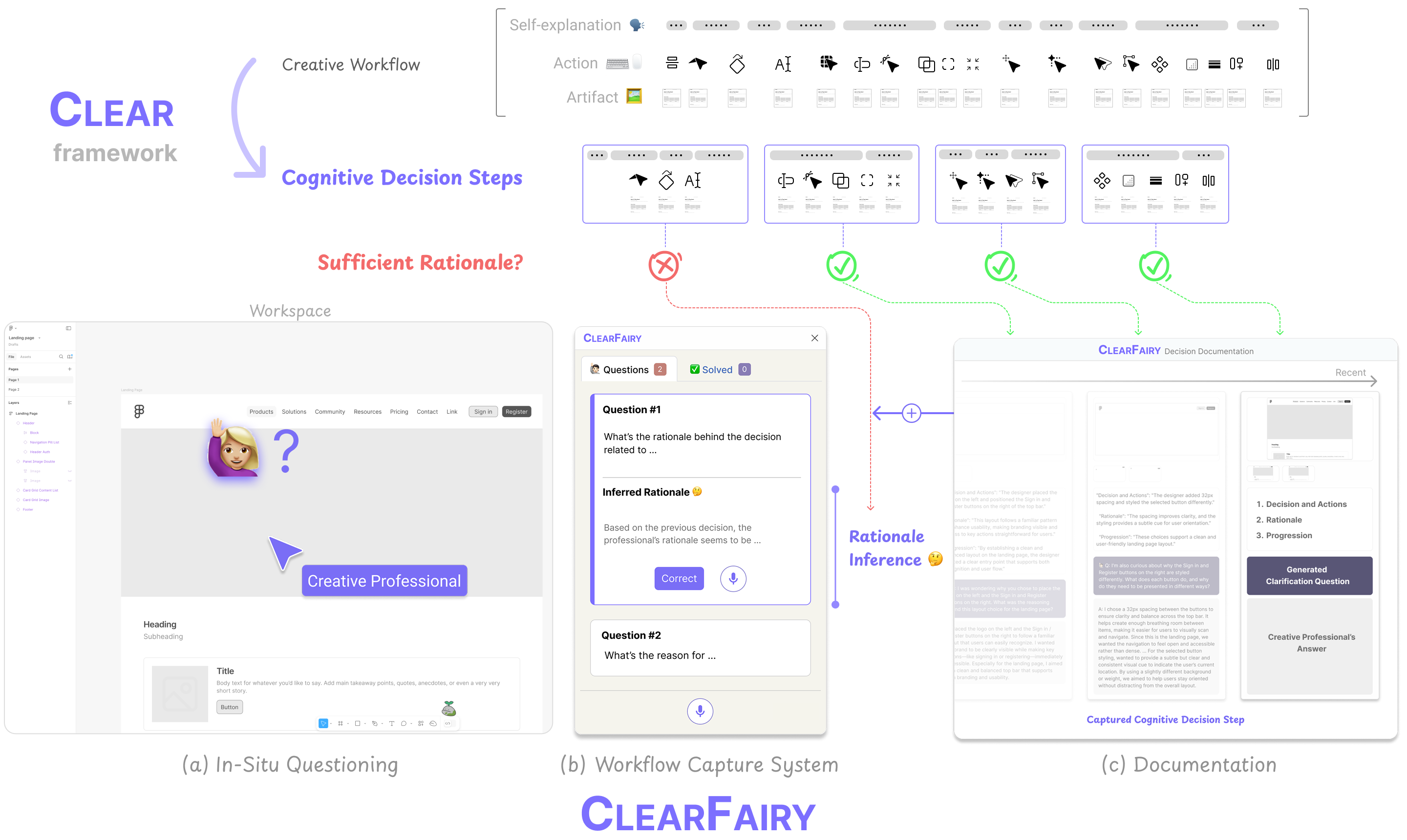}
  \caption{Overview of \approachname{} and \sysname{}. \approachname{} (top) is a workflow segmentation approach that structures workflows into cognitive decision steps by linking explanations, actions, and artifacts. Built on this approach, \sysname{} (bottom) captures these steps through an interactive pipeline: (a) In-Situ Questioning elicits rationales directly on the canvas, (b) the Workflow Capture System processes data via a Figma plugin, and (c) the Documentation page organizes the captured steps.}
  \Description{A conceptual diagram illustrating the CLEAR framework for capturing cognitive decision steps in creative workflows. The figure is divided into three sections: (a) In-Situ Questioning shows a Figma-like workspace where a creative professional works, with a floating label ``Creative Professional'' and a raised-hand emoji above a question mark. (b) Workflow Capture System displays a CLEARFairy sidebar UI with automatically generated questions such as ``What’s the rationale behind the decision related to...'' and an inferred rationale. The user can confirm or record their answer. (c) Documentation presents a structured interface showing captured cognitive decision steps, inferred rationales, and generated clarification questions under ``CLEARFairy Decision Documentation''. At the top, icons representing creative actions (e.g., selecting, aligning, editing) are annotated as ``Cognitive Decision Steps'' under the ``Creative Workflow''. Some actions are marked with green checkmarks (sufficient rationale), while one is marked with a red X (insufficient rationale), prompting rationale inference. Colored arrows connect the action timeline to the questioning interface and documentation, showing how decisions are clarified and recorded. Labels include ``Self-explanation'', ``Action'', ``Artifact'', and ``Rationale Inference''}
  \label{fig:teaser}
\end{teaserfigure}

% \received{20 February 2007}
% \received[revised]{12 March 2009}
% \received[accepted]{5 June 2009}

\maketitle

\section{Introduction}

Capturing and documenting professionals' cognitive processes and decision-making during creative workflows (e.g., design, writing, or video editing) is essential to understanding professional knowledge and applying this knowledge~\cite{fischer2020making}. As decisions in creative contexts often rely on heuristics and contextual insights, uncovering the underlying rationale is key in understanding each decision~\cite{lee1997designRationale}. Since much of creative work takes place in traceable software, professionals' decisions and rationales can be documented, where these documentations can support self-reflection~\cite{burge2000reasoning}, foster mutual understanding among collaborators~\cite{pena1997augmenting}, and enable knowledge transfer from experts to novices~\cite{kimball2000transfer}. Furthermore, these decisions can be collected as \textit{datasets} to analyze large-scale patterns~\cite{chang2020workflow, wang2018leveraging} or to train AI models that support creative workflows~\cite{fuchs2023modeling}.

To capture such decisions for sharing and reuse, prior approaches employ in-task annotations, retrospective interviews, and in-situ self-explanations. Among these, self-explanation---where professionals verbalize their rationales while working---has been shown to capture richer decision-making information than retrospective reflection~\cite{robert1995eliciting, collins1991cognitive}, while also imposing a lower burden than in-task annotations~\cite{krosnick2021Think}. This advantage is more pronounced in artifact creation workflows (e.g., 3D modeling or slide design) than in text production (e.g., programming) as the verbalization complements, rather than duplicates, the task's modality---aligning with Mayer’s Redundancy Principle~\cite{mayer2005cognitive}. Also, in creative domains such as UI/UX design, self-explanation is a familiar practice for professionals as they often prefer verbal explanations when documenting their decisions~\cite{feng2023understanding}.

However, self-explanation through verbalizing is not a natural part of everyday design work; rather, it is typically elicited in contexts such as critique, mentoring, or documentation. Within these contexts, capturing high-quality explanations through verbalization still faces important limitations. First, verbalized self-explanations are often incomplete or result in ``\textit{weak explanations}''~\cite{nielsen2002getting, ericsson1980verbal}---explanations that provide insufficient rationale for the decision---as opposed to \textit{strong explanations}. Second, while clarifying questions are important for capturing sufficient decision rationales, answering every question in real time can interfere with the creative process~\cite{mark2008theCost}. Third, verbal reports alone may not reveal implicit decisions~\cite{polanyi1968logic}, which often require contextual linkage to workflow traces (e.g., user actions or in-progress design artifacts) to surface subtle cognitive processes~\cite{nielsen2002getting}.

To address these challenges, we present \sysname{}, a think-aloud AI assistant for the UI/UX domain (\autoref{fig:teaser}). \sysname{} captures a professional's decision-making process while prompting sufficient rationale from the professional by automatically detecting insufficient rationale and generating lightweight clarifying questions. To achieve this, the system operationalizes \approachname{}, a workflow segmentation approach designed to support knowledge capture. Unlike static workflow segmentation defined through fixed rules such as sub-tasks or used tools, \approachname{} structures a creative professional's cognitive processes into ``\textit{cognitive decision steps}''---dynamic units of decision-making grounded in the real-time synthesis of linked \textit{self-explanations}, \textit{actions}, and \textit{artifacts}. For example, a single decision step may involve a professional noticing that a page title lacks visual emphasis, making focused visual adjustments, and explaining that the change clarifies hierarchy and identity. In another case, a professional might recognize that body text feels cramped, refining spacing or text size, and articulating how the change improves readability. By making cognitive processes traceable, this approach serves as the methodological foundation for \sysname{}, enabling the system to evaluate rationale sufficiency and elicit missing reasoning within each cognitive decision step.

Beyond simply prompting clarification questions, \sysname{} progressively learns from a professional’s previous Q\&As over the course of a knowledge-sharing session to anticipate and infer rationales for new questions, which the user can simply verify or refine. In doing so, the system reduces the burden of constant elaboration typically imposed by clarification practices, while still ensuring that decision rationales are captured in sufficient depth. \sysname{} is implemented as a Figma plugin\footnote{https://www.figma.com/}, embedding knowledge capture into the design environment itself. As shown in \autoref{fig:teaser}, it operationalizes the \approachname{} through a generative pipeline that segments decision steps, evaluates the sufficiency of their rationales, elicits or infers missing reasoning, and organizes the results into a structured workflow documentation. 

To evaluate the effectiveness of \sysname{} in capturing professionals' cognitive decision steps within knowledge-sharing scenarios of creative workflows, we conducted a within-subjects study with 12 creative professionals (i.e., interface design experts) across two web design tasks, where \sysname{} was compared to a baseline think-aloud approach. \sysname{} captured strong explanations for 83.2\% of all cognitive decision steps, compared to 13.9\% in the baseline. Notably, 85\% of the system-inferred rationales were accepted by participants, either as is or after minor revisions, reflecting strong alignment with their reasoning and easing the cognitive burden of decision justification. Participants reported that the system's inferences enabled them to recall deeper rationales, supporting the capture of implicit decisions---such as those based on expert conventions, aesthetic intuition, and personal preferences.

Furthermore, to illustrate the potential utility of the captured cognitive decision steps, we presented two exploratory applications: (1) inferring professionals’ next actions, and (2) simulating their design actions through a tool-calling agent to generate coherent design outcomes aligned with professional practices. For next-action inference, in 66.7\% of cases, professionals preferred the actions inferred by a Large Language Model (LLM) when it was provided with the cognitive steps, rather than without them. For design simulation, we built a ReAct-based~\cite{yao2023react} LLM-powered design agent in Figma that generates artifacts directly on the canvas. Providing captured decision steps to this agent produced outputs that were rated as more structured, detailed, and professionally aligned, compared to using only a task prompt.

For future research, we also release a dataset of 417 cognitive decision steps from 12 creative professionals, each containing a decision, associated actions, and rationale. Finally, we discuss broader implications of capturing implicit decision steps in creative workflows, focusing on their potential to support learning and advance AI agents.

This work presents the following contributions:
\begin{itemize}
    \item \textbf{\sysname{}}: An AI assistant for knowledge capture based on \approachname{} and implemented as a Figma plugin that elicits rationales behind decisions and progressively learns to infer the creative professional's rationale.
    \item \textbf{\approachname{}}: A new workflow segmentation approach that structures and captures \textit{``cognitive decision steps''} in creative workflows by dynamically linking action logs, self-explanations, and artifacts.
    \item \textbf{Empirical Findings from a User Study (N=12)}: Demonstrating that \sysname{} helps professionals articulate implicit decisions and capture significantly more strong explanations without increasing the burden of knowledge-sharing.
    \item \textbf{Cognitive Decision Steps with Reasoning Dataset}: A dataset that contains 417 cognitive decision steps captured from creative professionals' workflows, where each step includes a \textit{decision and actions}, \textit{contextual information}, and \textit{rationale behind the decision}.
\end{itemize}
\section{Related Work}
Our work builds on prior research across workflow segmentation, workflow capture tools, and generative model interventions to support creative professionals' reasoning and knowledge capture.

\subsection{Workflow Segmentation: An Approach Incorporating Rationale, Behavior, and Artifacts}
Workflow segmentation techniques have been proposed for diverse purposes, such as workflow navigation~\cite{pongnumkul2011pauseandplay, truong2021automatic, wang2021soloist}, key step search~\cite{pavel2014video, chang2017recipeScape, kong2012delta}, or authoring~\cite{chen2024tutoAI, chi2012mixT, lafreniere2014investigating, pu2023dilogics}.
To segment workflows, prior work has primarily relied on application usage logs (e.g., commands or tools)~\cite{lafreniere2014investigating, kong2012delta, chi2012mixT}, specific keywords in transcripts~\cite{truong2021automatic, chang2017recipeScape}, or predefined sub-goals~\cite{weir2015learnersourcing, pavel2014video, crowdsourcing2014kim}.
However, there is limited work that explores workflow segmentation that reflects the subject’s cognitive processes (e.g., knowledge and reasoning). While capturing logs or txranscripts separately provides partial information, existing computational methods lack the mechanism to integrate these multimodal signals to define meaningful decision boundaries.
Capturing a subject’s cognitive process in a workflow---why certain decisions were made \cite{brockmann2002tacit, lee1997designRationale}, what actions were taken \cite{droll2008seeing}, and what reasoning supports those decisions \cite{wickens1985engineering}---is crucial for knowledge sharing.
According to Lee~\cite{lee1997designRationale}, understanding a person’s cognitive process requires identifying the ``rationale'' that explains ``\textit{why}'' and ``\textit{how}'' a decision was made.
Prior work~\cite{carroll200315, moran2020overview} also argued that, beyond this rationale, it is crucial to capture the subject's behavior and produce artifacts to understand their implicit knowledge, reasoning, and decision-making process in detail.

In this work, we propose a workflow segmentation approach, called \approachname{}, that goes beyond simply recording multimodal data. Instead, it operationally synthesizes the links between self-    explanations, actions, and artifacts to dynamically structure the workflow. Unlike prior approaches that rely on fixed rules or single modalities, \approachname{} identifies the convergence of these three elements to define a ``\textit{Cognitive Decision Step}'' that reflects the user's thought process.

\subsection{Workflow Capturing and Documentation Tools}
Workflow capture has been actively studied not only for knowledge sharing but also for sharing collaborators’ decision-making processes~\cite{kim2021winder, leary2018charrette, park2018post, wang2020callisto} or enabling self-exploration of previously executed tasks~\cite{dragunov2005tasktracer}. 
For instance, Chronicle~\cite{grossman2010chronicle} and Screentrack~\cite{hu2019screenTrack} record the overall workflow and log the user's actions as a history. Waken~\cite{banovic2012waken} tracks user interactions with tools by reverse engineering them from task recording video.
To share decision-making processes among collaborators, Winder~\cite{kim2021winder} links user's explanations to workflow artifacts, and Charrette~\cite{leary2018charrette} allows users to save and share subsets of work artifacts as snapshots. 

Prior work~\cite{gruber1996generative, hertzum2009scrutinising} also suggests that self-explanation through think-aloud protocols can capture experts' reasoning without drastically interfering in their workflows.
Building on this, \textit{Co-notate}~\cite{rasmussen2019conotate} enables users to create video and audio recording to reflect on significant events (e.g., problems, ideas, or decisions) during the design process. 
Additionally, Krosnick et al.~\cite{krosnick2021Think} proposed \textit{think-aloud computing} that prompts users to share details during their workflows by visualizing the type and amount of explanation needed.
Similarly, \textit{creative livestreaming} has emerged as a widespread practice where professionals broadcast workflows and verbalize their process~\cite{fraser2019sharing}. While studies show that such streams contain rich, tacit knowledge~\cite{lu2018streamwiki}, utilizing this unstructured video content for systematic knowledge retrieval remains a challenge.

Crucially, the utility of such captured workflows depends on the quality of the accompanying rationale. Lee~\cite{lee1997designRationale} defines rationale as an explanation of ``\textit{how}'' and ``\textit{why}'' an artifact is designed, emphasizing that it must go beyond decisions to include justifications and alternatives. Building on this, we adopt the distinction between ``weak'' and ``strong'' explanations proposed by Gruber and Russell~\cite{gruber1996generative}, where a Strong Explanation is characterized by the presence of sufficient rationale. Drawing from prior work, a rationale is considered sufficient if it demonstrates causal reasoning by referencing situational relevant factors~\cite{guarino2009modeling}, drawing on prior knowledge or heuristics~\cite{brockmann2002tacit}, or engaging in comparative argumentation about alternatives~\cite{lee1997designRationale}. In contrast, a Weak Explanation relies on superficial justifications or personal preferences without such context. While self-explanation is a powerful method, ensuring the capture of sufficient rationale behind the decision remains a challenge in automated systems.

Although prior work has actively explored self-explanation-based workflow capture, these systems passively record user actions while relying on users to actively provide explanations. However, users may unintentionally overlook parts that should be explained~\cite{sela2000comparison}---often due to expert blind spots~\cite{wang2021seeing, nathan2003expert}---but continuous self-explanation is burdensome~\cite{van2003retrospective}. Therefore, our work presents a knowledge-capture tool that identifies points where rationales have been insufficiently explained in the user’s explanation and proactively prompts users with clarification questions.

\subsection{Generative Model's Intervention to Support Human's Thought Process}
As generative models, such as Large Language Models (LLMs) or Vision-Language Models (VLMs), are capable of producing contextually plausible answers or suggestions, prior work has investigated how users can leverage these responses effectively. In particular, generative models have been used to provoke student thinking in educational contexts~\cite{zhang2024using, favero2024enhancing}, prevent fixation and promote inspirational ideation~\cite{smith2025fuzzy, wadinambiarachchi2024theEffects, lee2024conversational, liu2024howAI, xu2024jamplate, pu2025ideasynth}, and, more broadly, facilitate human critical thinking~\cite{danry2023dont, mukherjee2023impactBot, lee2024one, drosos2025makes}. To interactively intervene in the human creative process, recent HCI research has proposed systems that offer real-time, context-aware suggestions to scaffold user reasoning and augment decision-making in various tasks~\cite {feng2024cocoa, riche2025ai, pu2025assistance, chen2024need, liu2024how}.
Beyond supporting creative and critical thinking, researchers have also explored employing LLMs as interactive partners that help users better articulate their thoughts by inducing reflection~\cite{wei2024leveraging}. For example, Thinking Assistants~\cite{park2023thinking} poses reflection-oriented inquiries grounded in expert knowledge, which helped users reach better decisions, and Kumar et al.~\cite{kumar2024supporting} showed that LLM-supported self-reflection increased self-confidence and improved learning outcomes in students. 

While generative models have shown promise in supporting the human thought processes through interventions, how such models can elicit users’ self-explanation during workflows remains underexplored. Specifically, think-aloud protocols could help externalize reasoning but place the cognitive burden of what to explain and when to users~\cite{nielsen2002getting, van1994think}, while generating clarification questions can interrupt the task workflow~\cite{mark2008theCost}. To address these challenges, we propose a system that leverages the reasoning capabilities of generative models to reduce this burden by predicting missing rationales and providing suggestions as knowledge capture progresses. Our system functions as a copilot-style assistant that supports rationale articulation, scaffolding users’ reasoning and helping make implicit knowledge more explicit during creative workflows.
\section{Design Goals}
Capturing creative professional's workflow through self-explanation faces three critical challenges: first, verbal reports alone often fail to reveal implicit decisions due to the lack of contextual linkage with workflow traces; second, verbalized explanations are frequently ``weak explanation'', providing insufficient rationale; and third, answering real-time clarification questions can disrupt the creative process. To overcome these limitations and effectively capture professionals' cognitive decision-making processes in their worklfow, we derive the following three design goals:

\textbf{DG1: Structure the professional’s workflow based on their cognitive decision-making process through their self-    explanations, action logs, and task artifacts.} A creative professional's expertise and knowledge are embedded in their decision-making process within the context of a task and are reflected through their \textbf{self-explanations}, \textbf{actions}, and their resulting \textbf{artifacts}~\cite{carroll200315, moran2020overview}.
Thus, we aim to capture data from all three channels and, by identifying interconnections between these, we can then structure and contextualize the professionals' overall decision-making process.

\textbf{DG2: Detect insufficient rationales in professionals' decisions and clarify reasoning through in-situ prompting.} Due to inherent blind spots~\cite{wang2021seeing, nathan2003expert} and the cognitive load associated with generating detailed self-explanations~\cite{nielsen2002getting}, experts tend to provide weak explanations~\cite{gruber1996generative} with insufficient rationale for their decisions.
While professionals can be asked to clarify after completing their workflows, this risks losing critical information regarding their decisions~\cite{danks1984Protocol}. To address this, we aim to identify decisions where an expert's rationale is insufficient and prompt clarifications from the user through context-aware, question-based interactions, while allowing them to respond at lightweight, opportune moments without disrupting their workflow.

\textbf{DG3: Infer the professional's rationale and document their decision steps to reduce the burden of knowledge-sharing.} While a professional engages in a workflow, asking them to self-explain and prompting them with clarification questions can potentially capture their cognitive processes and decisions. However, repeatedly answering such questions could impose a significant knowledge-sharing burden that can disrupt the creative workflow~\cite{van1994think, mark2008theCost}. To alleviate this burden, we aim to infer the rationale behind a professional's decision automatically based on their previously captured decision steps.

\section{\approachname{}: Cognitive Linking of Explanations, Actions, and Results for Workflow Segmentation}

To achieve DG1 (i.e., \textit{"structuring the professional's workflow based on cognitive decision-making"}), we propose \approachname{}, a new workflow segmentation approach in think-aloud-based workflows. As previously discussed, existing methods often fail to effectively capture the professional's full decision-making process in their workflow because they lack the contextual integration of user behaviors, rationales, and outcomes.

To bridge this gap, \approachname{} dynamically synthesizes self-explanations, actions, and resulting artifacts in real-time. This section details how our approach structures these signals into ``cognitive decision steps,'' defined as dynamic units of decision-making. These steps serve as the fundamental basis for our knowledge capture system to evaluate rationale sufficiency and trigger appropriate questions.

In designing this approach, we explicitly target a \textbf{Knowledge Sharing} context, recognizing that stakeholders have varying needs: novices require fine-grained steps to trace the expert's workflow in detail (including micro-decisions), while experts often prefer high-level summaries or key decisions for collaboration or self-reflection. To support this spectrum of needs, \approachname{} aims for \textit{lossless capture}, capturing every decision step as an ``atomic unit.'' Unlike approaches that rely on fixed sub-goals to segment workflows, our approach proposes ``adaptive boundaries'' that are constructed bottom-up by dynamically linking self-explanations, actions, and artifacts. This approach ensures that even fleeting micro-decisions are preserved, providing full traceability for novices to address expert blind spots while allowing for clustering or summarization for experts.

\label{framework}

\subsection{Definition of Cognitive Decision Step}

We define a \textbf{Cognitive Decision Step} as \textit{``a unit of decision-making defined based on \textbf{dynamic}, \textbf{adaptive}, and \textbf{contextual} boundaries, comprising one or more linked explanations, actions, and artifacts.''} This concept serves as the fundamental unit in our workflow segmentation approach and the operational basis for the rationale clarification-based knowledge capture system within the think-aloud process.

To accurately mirror the professional's fluid thought process, the boundaries of a decision step cannot be static. Instead, they must be (1) \textbf{dynamic} to capture decisions in real-time as they occur, (2) \textbf{adaptive} to accommodate individual variations in expertise and style, and (3) \textbf{contextual} to fully represent the decision by linking explanations with their corresponding actions and artifacts. Defining the segmentation unit with these properties is essential to structure the continuous workflow into discrete, meaningful cognitive units.

\textbf{Dynamic}: 
While prior work has explored methods for capturing decision-making units within users' workflows, they focused on \textit{retrospective} segmentation based on predetermined definitions of sub-goals in the workflow~\cite{weir2015learnersourcing, pavel2014video, crowdsourcing2014kim} or significant action types~\cite{lafreniere2014investigating, kong2012delta, chi2012mixT}.
However, a creative task workflow consists of diverse and sequential decisions, where capturing the rationale behind these decisions \textit{as they occur} is essential to gain a deeper understanding of the user's thought process~\cite{lee1997designRationale, regli2000aSurveyOfDesign}.
Thus, instead of post-hoc segmentation, \textbf{cognitive decision steps} are structured dynamically during the user's workflow in real-time.

\textbf{Adaptive}: 
Thought processes can vary across individuals based on their expertise levels and task execution styles---influencing the way that each individual makes decisions~\cite{sweller1988cognitive}. 
Due to these differences, a user's natural flow of thought in pursuit of one decision may vary not only across individuals but also depending on expertise and task context.
\begin{block}
    An experienced UI designer who wants to make a button stand out might make it bigger, adjust its contrast with the background, and add a drop shadow based on their knowledge of design trends. 
    On the other hand, a more novice designer with the same overall goal may first make it bigger to differentiate it from other buttons, adjust its contrast so it does not disappear in the background, and add a drop shadow to follow references they collected.
\end{block}
In the example, although the outcome achieved by both designers is the same, their thinking processes and individual decisions considered differ. Instead of fixed segments, each \textit{cognitive decision step} is flexibly bounded to capture variations in thought flow shaped by individual differences, expertise, and task context.

\textbf{Contextual}:
In a creative task workflow, each decision can only be fully understood through the professionals' explanations, actions, and the resulting artifacts~\cite{nielsen2002getting}. To understand a professional's cognitive decision steps within a workflow, it is necessary to observe these elements at a fine-grained level and identify the contextual links that bind them into decision steps. By examining these linked explanations, actions, and artifacts together, one can gain a deeper understanding of the professional’s decisions in context.

\begin{block}
    When editing a video, an editor might say, \textit{``This transition feels too abrupt---I'll add a fade and slightly slow down the clip before it.''}
\end{block}
Here, the editor's explanation provides insight into their thinking process but, to fully understand their cognitive decisions, we need more context about the actions taken by the editor (e.g., what type of fade was added, how much the clip slowed down) and the artifacts (e.g., what content is present in the previous and next clip). \textit{Cognitive decision steps} can represent the whole context by linking the explanations, actions, and artifacts, which are cognitively connected and relevant to a decision.

\subsection{Components of a Cognitive Decision Step}
\begin{figure}
    \centering
    \includegraphics[width=1.00\linewidth]{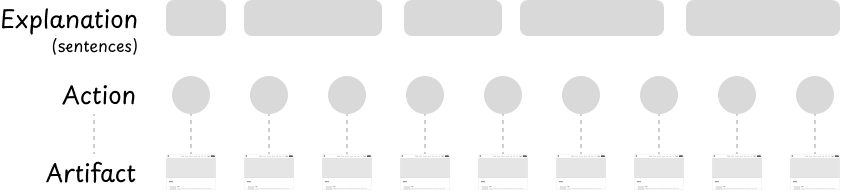}
    \caption{Three main components (Explanation, Action, Artifact) for constructing Cognitive Decision Steps in \approachname{}. Explanations provide the rationale behind several actions, and artifacts capture the resulting changes of each action.}
    \Description{This figure shows the structure of the components of \approachname{}, which connects three types of data to capture cognitive decision steps during creative workflows. At the top, sequences of self-explanation sentences reflect the user’s intent and reasoning. In the middle, discrete user actions represent interactions within the design environment. Each explanation contains one or more actions. At the bottom, artifacts, which are the snapshots of the evolving design, are tied to each action. By considering these data together, the \approachname{} segments the workflow into meaningful decision steps.}
    \label{fig:framework-component}
\end{figure}

In a self-explanation-based creative workflow capturing scenario, \approachname{} tracks three types of data: \textbf{self-explanations}, \textbf{actions}, and \textbf{artifacts} resulting from the professional's actions (\autoref{fig:framework-component}). Since we focus on creative workflows, such as creation or editing that produce an actual artifact, every action incurs changes in the artifact (\autoref{fig:framework-component}-Action and Artifact). Each component provides unique insights into the creative professional’s decision-making process:

\begin{itemize}
    \item \textbf{Explanations} reveal the professional’s reasoning and intent behind decisions.
    \item \textbf{Actions} show the digital interactions or operations performed within a workflow.
    \item \textbf{Artifacts} represents the visual state of the workspace after each action, recorded as screenshots that reflect the outcomes of those actions.
\end{itemize}

These components are structured into cognitive decision steps based on the following \textbf{principles}.

\subsection{Principles for Forming Cognitive Decision Steps}
This section defines how the three components can be linked: that is, how they can be considered cognitively connected and thus belong to the same cognitive decision step. Humans tend to speak and act in ways that reflect their ongoing thinking, leading to a coherent sequence of behavior. However, such cognitive relationships are often implicit and not directly observable. To address this, we propose objective rules for identifying cognitive links among low-level units of explanation, action, and artifact.

Building on these principles, we introduce an approach for structuring cognitive decision steps based on the natural continuity and interrelation between explanations, actions, and their resulting artifacts. Rather than imposing a predefined structure, our method identifies cognitive coherence through the following patterns:
\begin{itemize}
    \item \textbf{Cognitively Linked Explanations}: A sequence of explanations is directly linked in meaning and refers to the same object or sub-goal.
    \item \textbf{Cognitively Linked Actions and Their Resulting Artifacts}: Multiple actions occur sequentially and are logically connected in pursuit of a shared goal.
    \item \textbf{Cognitively Linked Explanations and Actions-Artifacts}: An explanation explicitly refers to a specific action or artifacts, or an action clearly reflects a preceding explanation.
\end{itemize}

\subsubsection{Cognitively Linked Explanations}
When a series of explanations flows continuously around the same object or intent, they often reflect a single cognitive unit in working memory rather than discrete steps, aligning with chunking theory in verbal protocols \cite{miller1956theMagical, ericsson1980verbal, danks1984Protocol, cowan2001theMagical}. In other words, uninterrupted segments of speech are likely already grouped based on a meaningful connection in the speaker’s mind, so capturing them as a single decision step also aligns with how our limited working memory naturally bundles linked information.

\begin{figure}
    \centering
    \includegraphics[width=1.00\linewidth]{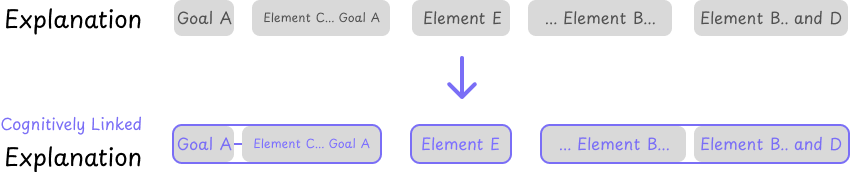}
    \caption{Construction of cognitively linked explanations. Cognitively linked explanations are created when consecutive utterances refer to the same goal or element within the workflow.}
    \Description{This figure shows how \approachname{} groups consecutive explanation segments based on their semantic and referential similarity. The top row shows individual explanation chunks mentioning various goals and elements. The bottom row demonstrates how these chunks are grouped into cognitively linked explanation clusters—each corresponding to a shared design goal or referenced element.}
    \label{fig:framework-sentence}
\end{figure}

We define cognitively linked explanations by identifying cases where consecutive sentences describe the \textbf{same mentioned purpose or element}, indicating that they originate from the same decision step (\autoref{fig:framework-sentence}). Even if some remarks are separated in time or interspersed with other thoughts, as long as they explicitly refer back to the same element or intent, they can be chunked into a single step in the timeline of decision-making.

\begin{block}
    For instance, if a designer consistently refers to ``making the button more visible'' across consecutive sentences, all such statements revolve around the same design goal and thus can be considered one cognitive unit (e.g., ``I want to make it bolder. Let’s change the color. Maybe also adjust the size a bit.''). 
\end{block}
\label{framework:linkedExplanation}

\subsubsection{Cognitively Linked Actions and Their Resulting Artifacts}
When multiple sequential actions take place under a sub-goal, they can form a ``chunked action'', an idea rooted in the extension of chunking from memory items to motor or procedural steps \cite{verwey1996buffer, lashley1951theProblem}. By grouping linked actions into one coherent unit, we reflect how people perform consecutive actions in a cognitive decision step.

\begin{figure}
    \centering
    \includegraphics[width=1.00\linewidth]{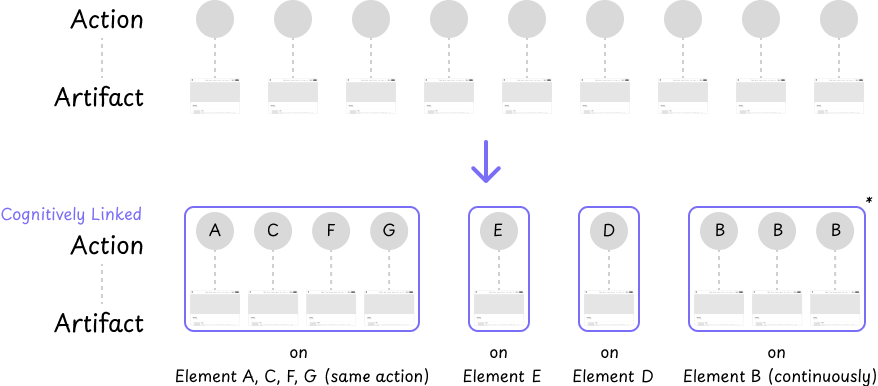}
    \caption{Construction of cognitively linked actions with artifacts. Cognitively linked actions are formed when the same action is performed on multiple elements, or when a sequence of actions is performed on the same element.}
    \Description{This figure shows how \approachname{} groups cognitively linked actions with artifacts. The top section shows individual actions and the corresponding artifacts they produce. In the bottom section, these actions are clustered into 4 groups: (1) one group contains the same action applied across multiple elements (A, C, F, G), (2) the other two groups capture a single action on a single element (D, E each), and (3) the other one with the repeated actions on a single element (B).}
    \label{fig:framework-actionArtifact}
\end{figure}

In particular, in a creative workflow, we define that linked actions can appear when various actions occur sequentially on one object or when the same action is performed on multiple objects (\autoref{fig:framework-actionArtifact}).

\begin{block}
    Consider a video editor is controlling hue adjustments on a single clip: she first shifts the greens toward yellow to warm up the foliage, then tweaks the reds to appear slightly more orange for skin tone correction. Though each change is a separate action, these are all parts of a single linked action aimed at adjusting the hue of the clip.

    The editor can also give a cross-dissolve transition effect to multiple clips at once to make the transition smooth. This kind of action should also be considered a linked action.
\end{block}
\label{framework:linkedActions}

\subsubsection{Cognitively Linked Explanations and Actions \& Artifacts}
After making links introduced in the previous steps, links between explanations and actions can be confirmed. If an explanation references a specific action or artifact---or an action clearly manifests a prior explanation---both are best treated as a single cognitive decision step, because explicit connections in discourse \cite{dijk1983strategies, kintsch1988theRole, clark1989contributing} or event structure \cite{zacks2007eventPerception, zacks2001eventStructure} naturally bind them into one conceptual unit. This dynamic linkage—whether through direct mention or bridging inferences—indicates that explanation and action with its results are cognitively inseparable, forming one decision step.

\begin{figure}
    \centering
    \includegraphics[width=1.00\linewidth]{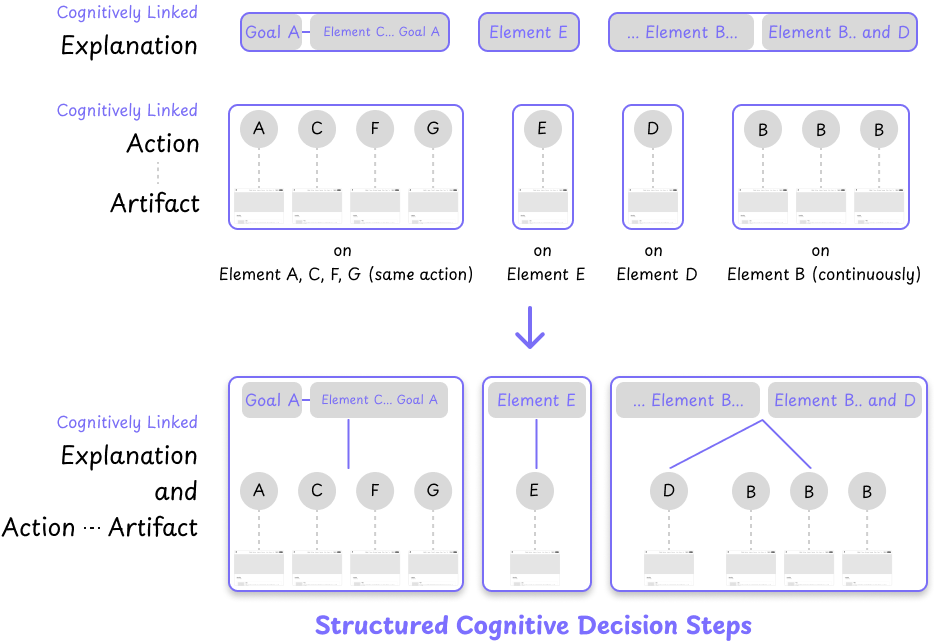}
    \caption{Formation of a Cognitive Decision Step by linking explanations with linked actions and artifacts. A cognitive link is established when an explanation directly refers to an associated action or artifact's part.}
    \Description{This figure shows how \approachname{} groups explanations, actions, and artifacts into structured Cognitive Decision Steps. The top row shows explanations grouped by shared goals or referenced elements, while the middle row displays groups of actions performed either on multiple elements using the same type of action or continuously on a single element, each paired with the resulting artifact. The bottom row integrates these layers: cognitively linked explanations are connected to their corresponding action-artifact clusters, forming complete cognitive decision steps.}
    \label{fig:framework-sentenceActionArtifact}
\end{figure}

We define cognitively linked explanations and actions (with their changes on the results) by identifying cases where an explanation explicitly describes an action before or after it is performed, as well as instances where an action is a direct response to an explanation (\autoref{fig:framework-sentenceActionArtifact}). These criteria establish a cognitive connection between what a user articulates and what they do, enabling a structured understanding of their decision-making process.

\label{framework:linkedExplanationAndActions}

\section{\sysname{}}
In this section, we introduce \sysname{} through a user scenario, followed by a detailed description of its pipeline.

\subsection{User Scenario and Interface}

With about five years of experience as an interface designer, \username{} is preparing materials to show junior designers at her company how she makes design choices and why. While working on her company's website landing page, \username{} realizes that this could be an ideal moment to explain to junior designers how to apply the company's design style and rules. Instead of separately working on these materials, she decides to use \sysname{} to document and explain her decisions while performing her own actual workflow.

\begin{figure}
    \centering
    \includegraphics[width=\linewidth]{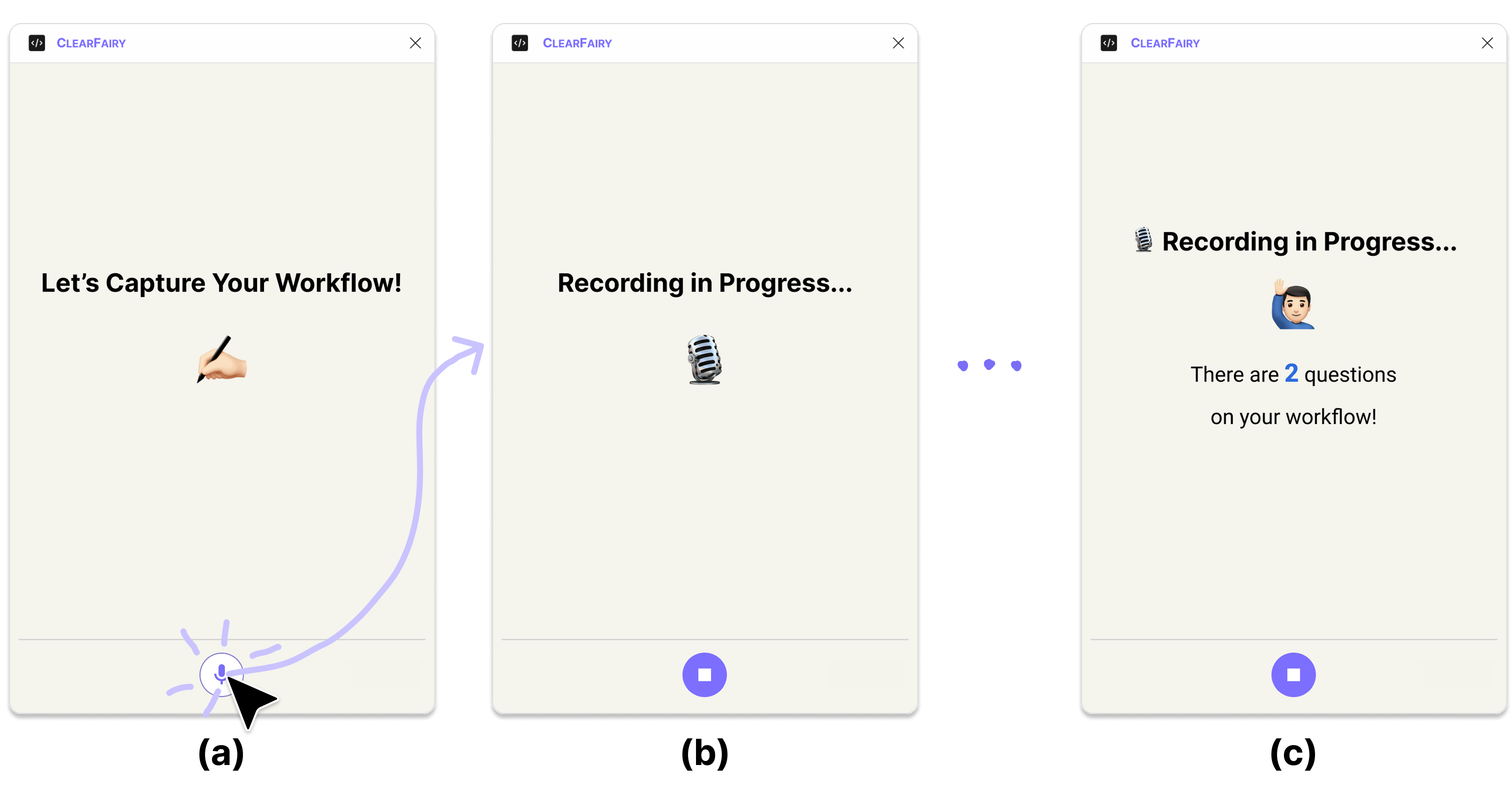}
    \caption{\sysname{} as a Figma plugin. Users can capture their workflow by clicking the recording icon, as shown in (a) and (b). (c) shows the interface when a clarification question is generated by \sysname{}.}
    \Description{This figure shows the user interface and the user flow for workflow capture in \sysname{}. In (a), the user starts recording by clicking the microphone icon, signaling the system to begin logging actions and capturing verbal explanations. (b) shows the interface during an active recording session, where the user narrates their work process. As \sysname{} analyzes the stream of actions and explanations in real time, (c) shows how it proactively detects points of weak or missing rationale and alerts the user with the generated clarification questions.}
    \label{fig:system-recording}
\end{figure}

\begin{figure}
    \centering
    \includegraphics[width=1.0\linewidth]{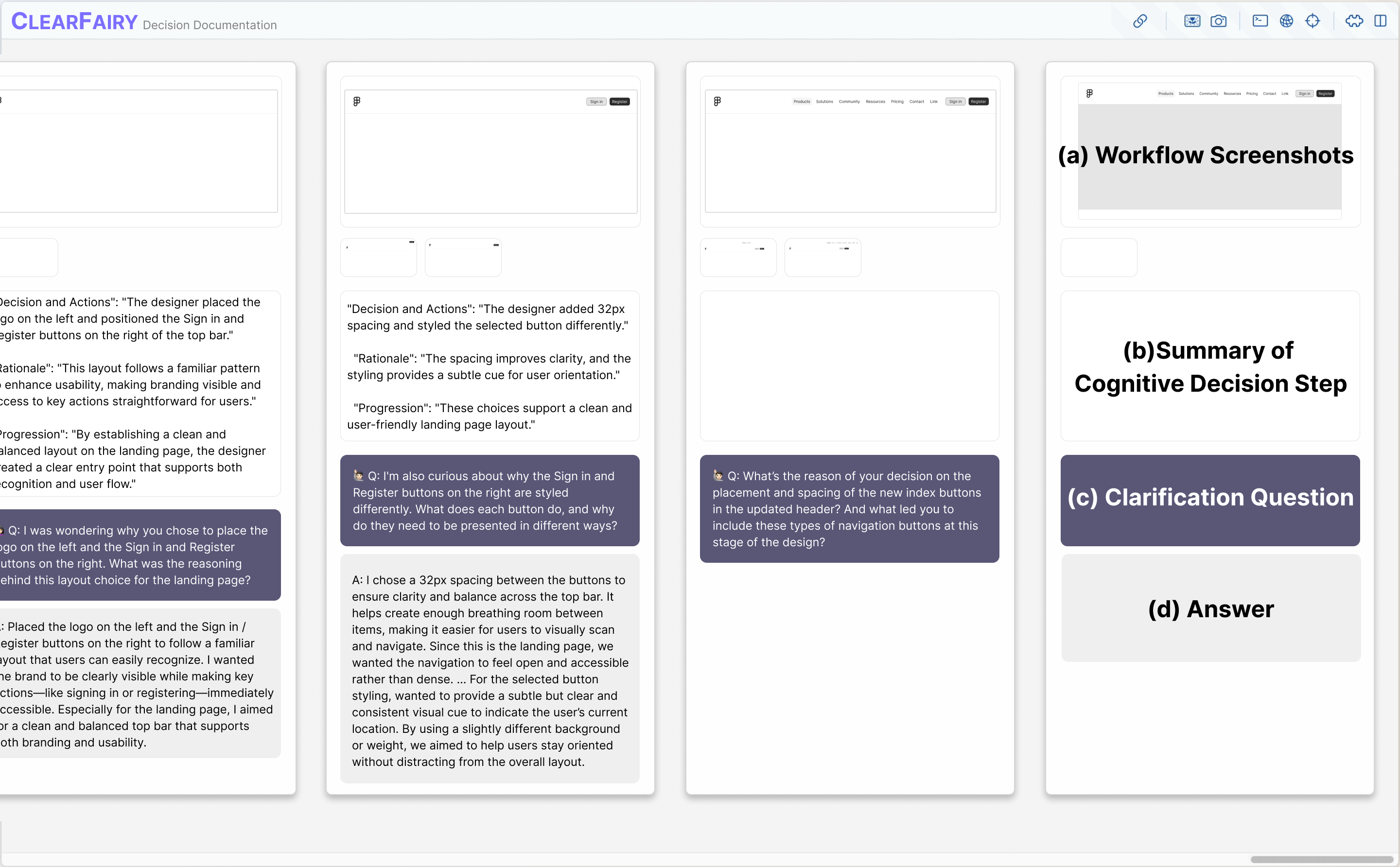}
    \caption{The documentation page for Cognitive Decision Steps in \sysname{}.}
    \Description{This figure presents \sysname{}’s decision documentation page, which shows key elements of a user’s design process into structured entries. For each cognitive decision step, the interface displays (a) visual snapshots, (b) a textual summary describing the action, rationale, and progression, (c) system-generated clarification questions for underspecified reasoning, and (d) the user's response to each question.}
    \label{fig:system-documentation}
\end{figure}

\sysname{} is implemented as a Figma plugin (\autoref{fig:system-recording}) and displays all documented decisions through a web interface (\autoref{fig:system-documentation}). As \username{} begins designing in Figma, she clicks the ``record'' icon button in \sysname{} and provides self-explanations of her design process in a think-aloud manner.

\begin{figure*}[!ht]
  \centering
  \includegraphics[width=\textwidth]{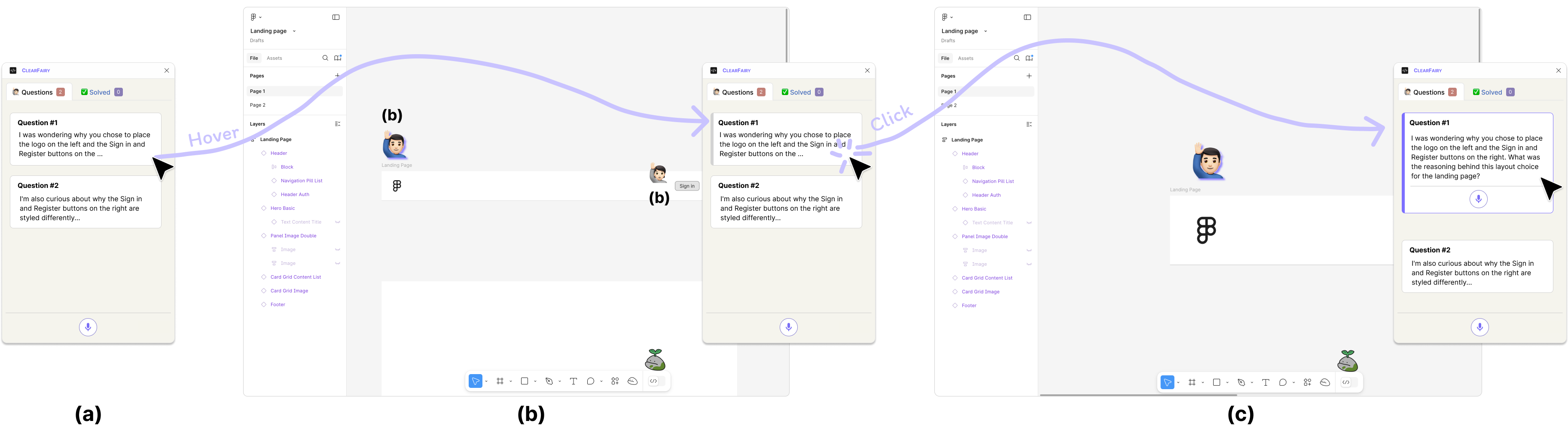}
  \caption{User interaction with clarification questions in \sysname{}. (a) shows a clarification question that appears after the user stops recording. When the user hovers over the question, a corresponding question mark is highlighted on the workspace, as shown in (b). Clicking the question mark in (b) moves the viewpoint to the location of the question, as shown in (c).}
  \Description{This figure shows the \sysname{}'s user flow of answering the clarification of weak or missing rationales during a design workflow. In (a), the system displays a list of clarification questions identified from the user’s previous explanations and actions. When the user hovers over a question, as shown in (b), \sysname{} highlights the corresponding design elements in the Figma design file to provide visual context. In (c), as a user clicks a question, the system shows it as a full prompt and activates a voice input option, allowing the user to provide additional rationale.}
  \label{fig:system-qna}
\end{figure*}

During the process of explaining her workflow while placing the necessary elements in the top bar of the landing page, \username{} sees an alert in \sysname{} with two questions about the current state of the work. She decides to finish designing the top bar first before answering the questions. Once complete, she clicks the ``recording stop'' button and then opens the questions (\autoref{fig:system-qna}-a). One question asked: ``I was wondering why you chose to place the logo on the left and the Sign in and Register buttons on the right. What was the reasoning behind this layout choice for the landing page?''
Another question asked: ``I’m also curious about why the Sign in and Register buttons on the right are styled differently.'' The design elements that the questions reference are marked (\autoref{fig:system-qna}-b), and when \username{} clicks on a question, \sysname{} adjusts the view to display the respective elements (\autoref{fig:system-qna}-c).

\username{} clicks the recording button for each question and answers these questions in detail, explaining the rationale for each design decision. For the first question about why the logo was placed on the left and the Sign in/Register buttons on the right, she explains that this arrangement follows a familiar convention and helps maintain a clean, balanced layout that makes key actions easy to locate. Then, \username{} returns to the design work and resumes the recording process.

\begin{figure}
    \centering
    \includegraphics[width=1.00\linewidth]{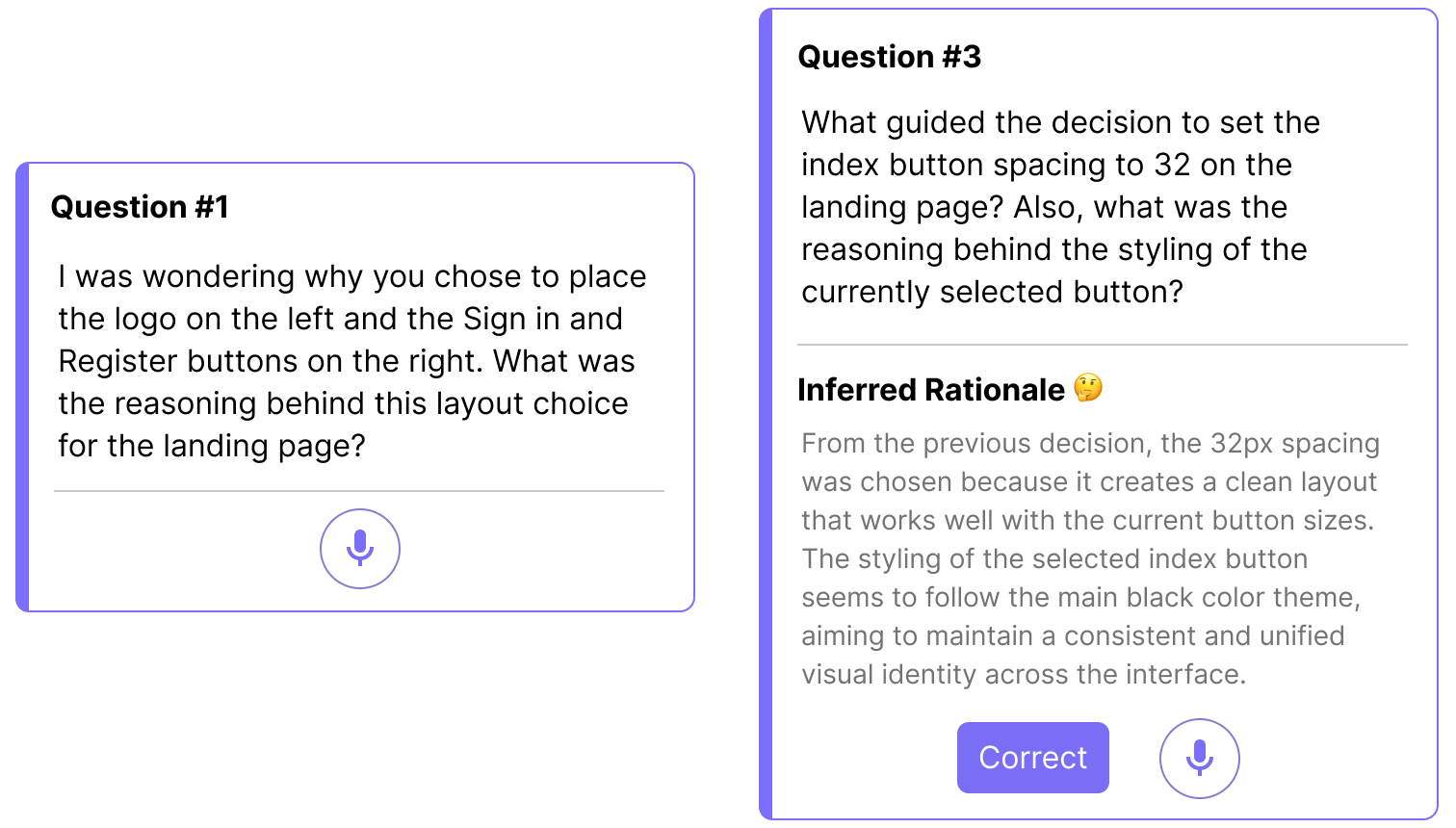}
    \caption{Rationale inference in \sysname{}. \sysname{} provides a rationale for the current decision based on the user's previous responses. The user can either accept the inferred rationale by clicking the ``Correct'' button or provide their own explanation using the microphone icon.}
    \Description{This figure shows two way of answering \sysname{}’s clarification question. On the left, users are asked a clarification question and prompted to verbally explain using the microphone button. On the right, the system proactively generates an inferred rationale based on prior explanations and actions for the next question. Users can confirm it by clicking "Correct" or offer further clarification via voice input. This can reduce the cognitive burden on users by allowing them to choose between elaborating their reasoning or validating suggestions made by the system.}
    \label{fig:system-rationaleInference}
\end{figure}

Once the top bar color and text button placement are finalized, \username{} again receives an alert in \sysname{} indicating there are two new questions. After pausing the recording and checking the questions, \username{} sees that \sysname{} now shows its own inferred rationale for why she may have made those design decisions, alongside the questions (\autoref{fig:system-rationaleInference}). For example, \sysname{} infers that the index button spacing of 32px was chosen to create a clean layout that works well with the current button sizes, and that the styling of the selected index button follows the main black color theme to maintain a consistent visual identity. \username{} judges one of these inferred rationales to be correct and clicks the ``Correct'' button. For the other question, she feels she could offer additional reasons for the decision, so \username{} directly answers it.

\sysname{} also provides a documentation webpage that consolidates the captured design process (\autoref{fig:system-documentation}). This page organizes each decision together with its rationale, the system’s clarifying questions, and the user’s answers, ensuring that the reasoning is preserved alongside the design artifacts. By continuously updating this record as \username{} progresses through the landing page, \sysname{} creates a living documentation of the workflow that can be revisited, shared, or reused later.

\subsection{Technical Pipeline}

\begin{figure*}[!ht]
  \centering
  \includegraphics[width=\textwidth]{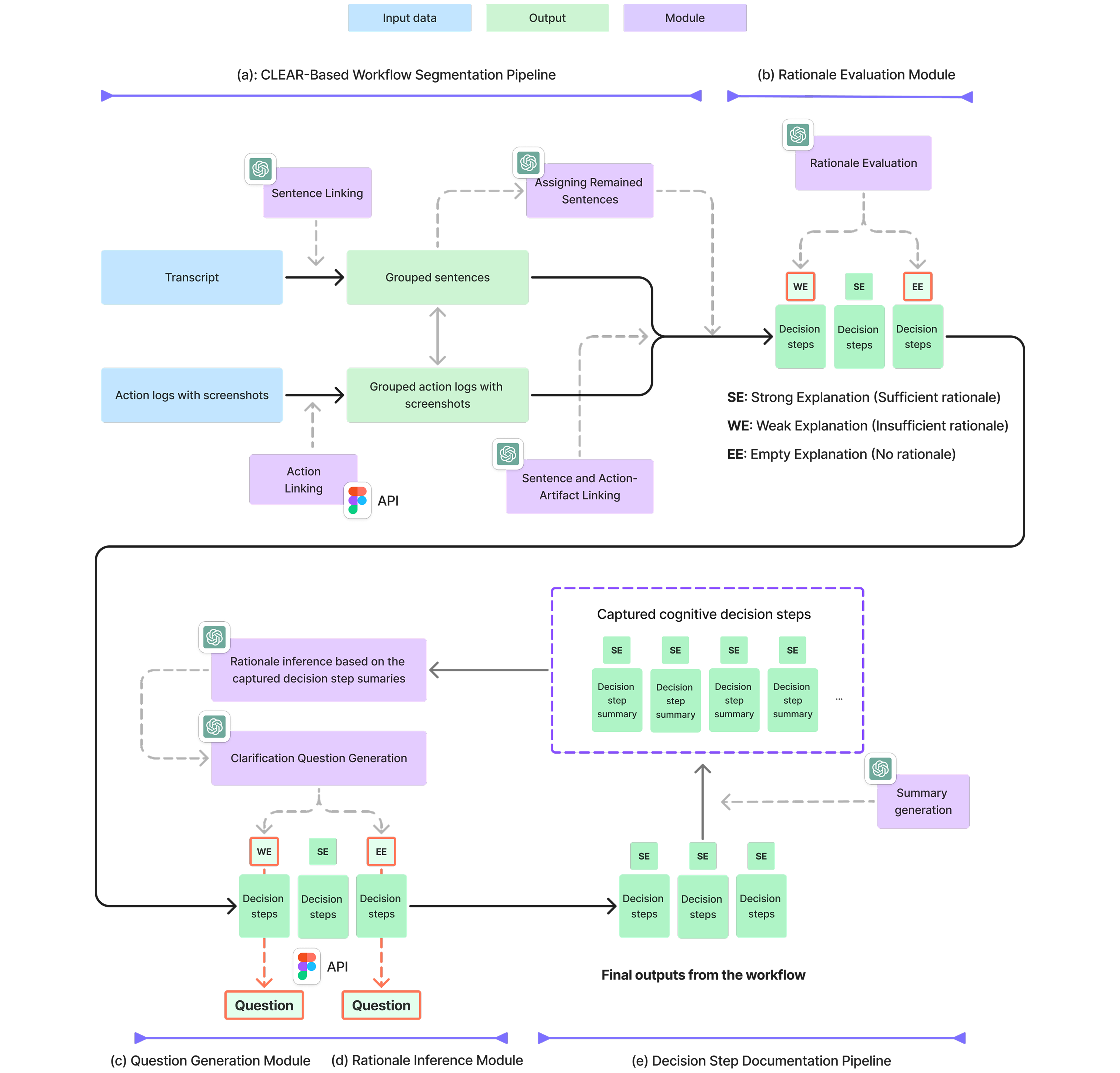}
  \caption{Technical pipeline of \sysname{}. (a) \approachname{}-based workflow segmentation pipeline, (b) rationale evaluation module, (c) question generation module, (d) rationale inference module, and (e) decision step documentation pipeline.}
  \Description{This figure shows the architecture of \sysname{}’s technical pipeline for transforming the raw design workflow data into structured cognitive decision steps. The pipeline begins with input data: transcripts and action logs with screenshots. These are processed through the \approachname{}-based segmentation module (a), which groups sentences and actions, then links them into cognitive decision steps. In module (b), the rationale evaluation component categorizes each step as having a strong (SE), weak (WE), or empty (EE) explanation. In the question generation module (c), it uses an LLM to produce clarification prompts for each decision step labeled as weak or empty. In (d), a rationale inference module attempts to infer missing reasoning based on prior user responses and contextual decision history. Finally, in module (e), the system generates summaries of well-reasoned steps to create a dataset of structured cognitive decision steps.}
  \label{fig:pipeline}
\end{figure*}

In this section, we describe the technical pipelines of \sysname{}. 
The technical pipeline consists of five interconnected components, each supporting a key stage in the real-time structuring of cognitive decision steps. As illustrated in Figure~\ref{fig:pipeline}, the system begins by segmenting the user’s workflow into cognitive decision steps. It then evaluates the sufficiency of the rationale for each step and, when needed, prompts clarification or infers missing reasoning. Finally, it documents the structured decision steps into a searchable and reusable format.\\

The five components of the pipeline are:

\textbf{\approachname{}-Based Workflow Segmentation Pipeline}: Automatically segment a creative professional's workflow into their cognitive decision steps (\autoref{fig:pipeline}-a)

\textbf{Rationale Evaluation Module}: Detect cases where the explanation provided by the user in a cognitive decision step either contains insufficient rationale (a weak explanation) or provides no rationale at all (\autoref{fig:pipeline}-b).

\textbf{Question Generation Module}: In the case of decision steps that have a weak or empty rationale, generate and prompt rationale clarification questions (\autoref{fig:pipeline}-c).

\textbf{Rationale Inference Module}: Based on the decision steps' rationale previously provided by the user, infer the user's rationale for a new decision step that has a weak or empty explanation (\autoref{fig:pipeline}-d).

\textbf{Decision Step Documentation Pipeline}: This pipeline (\autoref{fig:pipeline}-e) documents the user’s workflow by summarizing (Decision Step Summary Module) and organizing their cognitive decision steps, including:
\begin{itemize}
    \item Decision and Actions: the core design decisions and actions (with their details) taken by the user in this step.
    \item Rationale: A detailed reason behind the user's key decision.
    \item Progression: A detailed description of how the workflow logically progressed based on the user's decisions.
    \item Screenshots of Artifact: Visual records of the working artifact.
    \item Rationale Clarification Question and Answer: If there was a Q\&A process in this step.
\end{itemize}

Each part employs various Large-Language Model- (LLM) and Vision-Language Model- (VLM) based prompting methods.

\subsubsection{Workflow Segmentation into User’s Cognitive Decision Steps Based on \approachname{}}
\label{system:workflowSegmentation}
Based on the proposed \approachname{} in Section \ref{framework}, we introduce a method for segmenting a design workflow into the cognitive decision steps using the user’s think-aloud explanations, action logs, and screenshots of the artifact.

To segment workflows into cognitive decision steps, we follow a four-stage process designed to mirror how reasoning unfolds in context. We first group related explanations and actions independently, then link them based on semantic and visual context. Finally, we assign any leftover statements to ensure coherence.

The pipeline consists of four modules that leverage LLM and VLM: 1) sentence grouping (Sentence Linking Module), 2) action grouping (Action Linking Module), 3) linking between sentences and actions (Sentence and Action \& Artifact Change Linking Module), and 4) assigning remaining sentences (Sentence Assigning Module).

\paragraph{\textbf{Explanation Sentence Linking Module}}
The user’s spoken explanations are transcribed with STT (Google Cloud Speech-To-Text\footnote{https://cloud.google.com/speech-to-text?hl=en}) and segmented into sentences using the KSS (Korean Sentence Splitter) library.

To find linked sentences, we designed a few-shot prompt (Appendix \ref{appendix:prompt_sentence_linking}) based on \approachname{} (Section \ref{framework:linkedExplanation}) which groups consecutive sentences that describe the same: 1) design element (e.g., font, layout, button), 2) properties (e.g., size, color, shape), 3) design purpose, or 4) a specific design sub-task.

\paragraph{\textbf{Action Linking Module}}
Based on Section \ref{framework:linkedActions}, we defined the rules for linking related actions when a user 1) performs the same action consecutively on a single element or 2) takes the same actions on multiple elements simultaneously. In Figma, we can track the user’s action log whenever an action event occurs by using the Figma Plugin API \footnote{https://www.figma.com/plugin-docs/}. 

The tracked action log stores the ID of the element on which the action was performed and details related to the changes made to that element (e.g., the type of property changed and the specifics of the change). For each action log, we also capture a screenshot of the entire Figma canvas and store it together with the log.

However, even if actions are performed consecutively on a single element, they may have different purposes. For example, adjusting the size of a rectangular button and changing its color could serve different purposes, whereas resizing a rectangular button and moving its position might serve the same layout-related purpose. This indicates that there are various ways to group actions performed on a single element. We compared three methods of linking actions on the same element and selected the approach that groups only sequential identical actions on the same element (see Section \ref{techEval:segmentationResults}).

\paragraph{\textbf{Sentence and Action \& Artifact Linking Module}}
After making linked actions and linked sentences, establishing cognitive steps requires linking the actions with the sentences by identifying their relationships as we described in \approachname{} approach (Section \ref{framework:linkedExplanationAndActions}). To create these links, we designed a zero-shot VLM-based prompt (Appendix \ref{appendix:prompt_sentence_action-artifact}) that finds connections between action-screenshot pairs and sentences.

To perform this task effectively, it is essential to understand the design context. Rather than simply providing action logs and sentences, we include the screenshots as input so that the VLM can better comprehend the design context when creating the links. Specifically, we designed a prompt that links each sentence group to the relevant timestamp(s) in the action logs and their corresponding artifact screenshots whenever the transcript describes or relates to the design action or the artifact’s screenshot. Through this process, segments composed of explanations, actions, and artifacts are created. 

\paragraph{\textbf{Sentence Assigning Module}}
Since the segments created in the previous linking process are formed solely based on clear and objective relationships between sentences and between sentences and actions, certain types of explanations---such as evaluations or plans (``Looks fine'' or ``Now, I'm planning to revise this part a bit further'') that might emerge in a think-aloud environment \cite{krosnick2021Think}---could remain unlinked. To find and decide whether to include these remaining sentences in the most relevant segment or not, we designed another LLM-based assignment module (Appendix \ref{appendix:setence_assigning}).

In this module, we first gather any sentences not assigned to any segment. For each missing sentence, we prompt the LLM with: (1) the full transcript for broader context; (2) the unassigned sentence itself; (3) the immediately preceding left segment, which is identified by the grouped sentence number in the sentence grouping module; and (4) the immediately following right segment. We instructed the module to choose whether the unassigned sentence should attach to the left segment, the right segment, or be left alone as ``unrelated'', based on semantic continuity and additional instructions (e.g., if it describes an evaluation of a just-finished task, assign left; if it proposes a plan for a next step, assign right). This ensures a more natural grouping of any remaining think-aloud statements.

\subsubsection{Rationale Evaluation Module on Cognitive Decision Steps}
\label{system:rationaleEvaluationModule}
To check whether each user’s decision steps contain a sufficient rationale for the decision or not, we designed an LLM-based rationale evaluation module. A rationale is an explanation of how and why an artifact, or some portion of it, is designed the way it is \cite{regli2000aSurveyOfDesign}. Rationale offers more than the decisions; it also offers the reasons behind each decision, including its justification, other alternatives considered, and argumentation leading to the decision \cite{lee1997designRationale}.

Building upon the definition of weak and strong explanation by Gruber and Russell \cite{gruber1996generative}, we defined three types of explanation quality: sufficient rationale (``Strong explanation''), insufficient rationale (``Weak explanation''), and empty rationale (``Empty explanation''). Specifically, we defined a specific category within type of explanation based on the prior work.

A strong explanation clearly demonstrates how and why a decision was made, often by 1) referencing situational relevant factors (mark as S-SR) \cite{guarino2009modeling}, 2) drawing on prior knowledge, principles, or heuristics (S-PK) \cite{brockmann2002tacit}, or 3) considering alternatives with concrete reasoning on the comparison dimensions (S-CA) \cite{lee1997designRationale}. 
A weak explanation \cite{gruber1996generative} depends on superficial justification on the situation relevance (W-SR), personal preference \& heuristics without referencing relevant prior experience (W-PK), or lack of decision factors even when alternatives are mentioned (W-CA). Finally, an empty explanation provides little to no reasoning (E). These categories reflect the essential details that a sufficient rationale must include.

We prompt the few-shot LLM-based module (Appendix \ref{appendix:prompt_rationale_evaluation}) by providing (1) the definitions of each explanation category (Strong/ Weak/Empty) and (2) a single cognitive decision-step excerpt from the transcript. We instruct the module to identify which of the predefined categories (e.g., S-SR, S-PK, S-CA, W-SR, W-PK, W-CA, E) apply to the designer’s explanation, along with a short rationale for why those categories are chosen. After obtaining the model's output (e.g., {S-SR, W-CA}), the code groups the categories as ``Strong'', ``Weak'', or ``Empty''. It then applies simple logic to decide on an overarching type: if both ``Strong' and ``Weak'' are present, it becomes ``strong''; if only ``Weak'' is present, it becomes ``weak''; if Empty is present alone, it becomes ``empty'', and so on.

\subsubsection{Question Generation Module}
We designed a zero-shot LLM-based module (Appendix \ref{appendix:prompt_question_generation}) that generates a clarification question to prompt the users when the Rationale Evaluation Module's (Appendix \ref{appendix:prompt_rationale_evaluation}) output is ``Weak'' or ``Empty'' explanation. In this prompt, we provide the definitions and criteria like in Appendix \ref{appendix:prompt_rationale_evaluation} and the Rationale Evaluation together. Finally, the prompt requests a specific, context-based follow-up question that could address the explanation’s shortcomings and help transform it into a ``Strong Explanation''.

The generated questions are placed on the relevant design objects in Figma by calculating the object’s location in the decision step's action logs (\autoref{fig:system-qna}-c). By localizing the questions on their workspace directly, the user can notice it without moving to a separate screen, and they can easily distinguish what parts require more explanation to provide a sufficient rationale.

\subsubsection{Decision Step Summary Module}
To document the decision step---when a decision step with a ``strong explanation'' is captured or when the user responds to a clarification question---we designed an VLM-based module (Appendix \ref{appendix:prompt_decision_step_summary_generation} and \ref{appendix:prompt_decision_step_summary_generation_with_answer}) that can summarize the relevant decision step. We prompt the model with the transcript containing that decision step, the specific decision step (excerpted from the overall transcript), and screenshots of the artifacts included in the cognitive decision step. The model is instructed to summarize the main design decisions, actions, rationale, and progression involved in the decision step. The output is then stored in the database, displayed in the web system, and used as input for the rationale inference module.

\subsubsection{Rationale Inference Module}
To lower the burden of responding to questions, we designed an LLM-based rationale inference module (Appendix \ref{appendix:prompt_rationale_inference}) that infers a missing or insufficient rationale for a specific decision step by referencing the user’s past, well-justified decisions. We provide the full transcript and the specific decision step, similar to the decision step summary module, and the reasons why the rationale evaluation module classified it under a weak or empty category, including the documented previous decision step summaries.

We instruct the module to (1) review the full context and pinpoint the decision step that lacks rationale, (2) examine previous decision summaries for overlap in specific elements, actions, or design goals, and (3) derive a plausible rationale based on any concretely similar past decisions, rather than abstract parallels or general design principles.

\subsection{Implementation Details}
\sysname{} is implemented as a Figma plugin based on the ReactJS framework for front-end and a Python Flask server with SQLite as the back-end. The LLM and VLM model in every module is OpenAI's GPT-4o (\texttt{gpt-4o-2024-08-16}) with a temperature of 0.3. Our pipeline performs real-time segmentation when thresholds are reached—for example, more than 20 action logs, more than 20 recognized sentences, or a pause longer than three seconds since the last action.
\section{Technical Evaluation}
In this section, we evaluate our workflow segmentation pipeline (Section \ref{system:workflowSegmentation}) and rationale evaluation module (Section \ref{system:rationaleEvaluationModule}). We scoped the technical evaluation to these modules because they require validation of computational accuracy, whereas the effectiveness of interactive components (i.e., Question Generation and Rationale Inference modules) is best assessed through actual usage in the user study (Section \ref{evaluation}).

We conducted a technical evaluation through a data collection study. In the study, we collected three types of workflow data from six interface designers: 1) think-aloud recordings, 2) action log data, and 3) screenshots of the entire content in the design file. Then, we processed the collected data to create ground-truth dataset of cognitive decision steps and rationale evaluations for each decision step, and then evaluated our pipeline's output on this dataset.

\subsection{Data Collection Study}
To capture the self-explanations, action logs, and screenshots, we invited designers to use our system without the question generation and rationale inference features. 
% When recording starts, the plugin captures the three types of data. The plugin takes a screenshot of the entire Figma canvas only when an action is performed.

\subsubsection{Participants}
The study involved six professionals (4 females and 2 males; Age\_MEAN=25.5, SD=1.2) who were compensated at a rate of 50,000 KRW/hour for participation. Every participant has an educational background in design (e.g., major in graphic design, industrial design, or interface design) and design experience of more than 1 year in their field (MEAN=2.9 years and SD=0.6).

\subsubsection{Procedure}
The data collection process was structured into two phases that took around 60 minutes in total: an introduction phase (15 mins) and a design task phase (40 mins). Participants were instructed to imagine themselves as a live streamer who is demonstrating and explaining their design process to a live audience. 
To minimize potential priming effects from the example videos shown for context, we explicitly instructed participants to explain their process in their own unique style. They performed a web portfolio design task in Figma, which consisted of two sub-tasks: designing the main page and the project page for a design portfolio website. During each task, participants were instructed to start the system's recording when they started a design step and stop the recording when said step was completed. During recording, participants explained their thought process following the think-aloud protocol~\cite{ericsson1980verbal}.

\subsubsection{Statistics of the Collected Data}
A total of 52 recordings were collected, excluding three recordings due to excessively long action logs that surpassed the gpt-4o-2024-08-16 model's max input token length. The average length of a recording was 205.90 seconds (SD=65.43, MIN=99, MAX=313), resulting in a total of 10,707 seconds of recorded data. The average number of action logs per recording was 36.81 (SD=12.53, MIN=13, MAX=67), bringing the total to 1,914 actions. The average number of sentences in the transcribed audio was 14.85 per recording (SD=9.75, MIN=4, MAX=47), totaling 772 sentences across all recordings.

\subsubsection{Creating the Ground-Truth Dataset}
To create the ground-truth dataset, two authors watched through the recordings and segmented them into decision steps. Specifically, the authors focused on segmenting the transcripts of the audio recordings. The two authors first segmented the transcripts for 12 recordings (23\% of the total) individually and then compared the results. The inter-rater reliability (IRR) based on Cohen's Kappa was 0.672, indicating substantial agreement. Any conflicts in the segmentations of the initial 12 recordings were merged. Afterward, the two authors each segmented 20 more videos, creating a ground-truth dataset of transcripts segmented into decision steps for 52 recordings.

\subsection{Performance of Cognitive Decision Step Segmentation Pipeline}
\label{techEval:segmentation}
We verified the design of our segmentation pipeline by conducting an ablation study where we tested the effect of each module in Section~\ref{system:workflowSegmentation}. We compared the results with a baseline, which segmented the recordings based on only the transcript data---adopting the segmentation prompt from TutoAI~\cite{chen2024tutoAI}. The baseline instructs an LLM to segment the transcript at points where the user makes distinct decisions by considering the relationship between sentences in the transcript.
All modules in our pipeline and the baseline used gpt-4o-2024-08-16 through the OpenAI API with the temperature set to 0.0.

\subsubsection{Metrics}

To evaluate the performance of segmentation, we calculated the precision and recall of the predicted transcript segmentations. Also, we utilized the WindowDiff metric~\cite{pevzner2002aCrique}, which measures the difference between two texts by comparing the lengths of matching and non-matching segments in a sliding window to account for minor segmentation misalignments.

\subsubsection{Results}
\label{techEval:segmentationResults}
\paragraph{\textbf{According to the approach for grouping actions}} For our Action Linking Module (Section~\ref{system:workflowSegmentation}), we considered that there could be multiple distinct methods to group actions performed on a single element in Figma. Rather than deciding on a single method from the start, we explored how performance at segmenting cognitive decision steps depends on how we link the actions on an element. Specifically, we defined the following three versions of action grouping:
\begin{itemize}
    \item V1: Group all consecutive actions on a single element, regardless of action type.
    \item V2: Group consecutive actions if they are all \textit{layout-related} (i.e., changing shape or position) or if they are all \textit{style-related} (i.e., any action not related to layout).
    \item V3: Group only consecutive identical actions.
\end{itemize}

We applied the same sentence grouping and assigning modules in all cases, only varying the action grouping method to generate cognitive decision steps, and then evaluated the results (\autoref{tab:techEval-action}).

Although V3 has the lowest precision among the three (V1: 0.7655, V2: 0.7583, V3: 0.7446), it achieves the highest recall (V1: 0.6696, V2: 0.7228, V3: 0.7532) and the highest F1 score (V1: 0.6841, V2: 0.6947, V3: 0.7211). In terms of WindowDiff, V1 and V3 show similar results, indicating that the predicted segmentation for both methods is highly similar to the ground truth. Although V1 shows overall good performance, it groups too many actions together, causing it to segment accurate decision steps too broadly and, consequently, perform lower on recall. 

\begin{table}[]
\begin{tabular}{lrrrr}
\hline
% \rowcolor[HTML]{EFEFEF} 
Condition & \multicolumn{1}{l}{WindowDiff} & \multicolumn{1}{l}{Precision} & \multicolumn{1}{l}{Recall} & \multicolumn{1}{l}{F1} \\ \hline
GAv1\_GS\_SA & \textbf{0.4668} & \textbf{0.7655} & 0.6696 & 0.6841 \\ 
GAv2\_GS\_SA & 0.5011 & 0.7583 & 0.7228 & 0.6976 \\
\textbf{GAv3\_GS\_SA} & 0.4712 & 0.7446 & 0.7532 & \textbf{0.7211} \\ \hline
BASELINE & 0.7624 & 0.5308 & \textbf{0.9270} & 0.6460 \\ \hline
\end{tabular}
\caption{Segmentation performance according to the method for grouping actions. ``GA'' refers ``Group Actions'' with ``v1'', ``v2'', and ``v3'' referring to the method. ``GS'' refers to ``Group Sentences'' and ``SA'' to ``Sentence Assigning''. Best performance per column is in \textbf{bold}.}
\Description{This table compares segmentation performance across different action grouping strategies. All variants include sentence grouping (GS) and sentence assigning (SA), while GAv1, GAv2, and GAv3 differ in how actions are grouped. GAv3_GS_SA achieves the best overall F1 score (0.7211) and a good balance between precision and recall. GAv1_GS_SA yields the lowest WindowDiff and highest precision. The BASELINE method, which uses only transcript data, achieves the highest recall but performs worst in precision and segmentation accuracy.}
\label{tab:techEval-action}
\end{table}

In contrast, the baseline shows the lowest performance. Specifically, the baseline oversegments the transcript---leading it to achieve high recall as it predicts a large number of segment boundaries, but low recall as it subdivides the transcript more than the ground-truth. The baseline also has the highest WindowDiff value further indicating that the segmentation is not accurate. Based on these results, we decided that our pipeline will perform action grouping in the same approach as V3.

\paragraph{\textbf{With or without the sentence grouping module}} \autoref{tab:techEval-sentenceLinking} demonstrates that sentence grouping (Section~\ref{system:workflowSegmentation}-Sentence Linking) is necessary to achieve high performance when segmenting a workflow into cognitive decision steps. 
Specifically, we compare when the sentence grouping model is used (GAv3\_GS\_SA) to when it is not used (GAv3\_IS\_SA). We observe that, without sentence grouping, the WindowDiff value increases (GAv3\_GS\_SA: 0.4712; GAv3\_IS\_SA: 0.5955) and the precision decreases (GAv3\_GS\_SA: 0.7446; GAv3\_IS\_SA: 0.5851). In other words, if mutually relevant sentences are not linked first, the pipeline ends up generating many segments, which leads to inaccurate results. Consequently, although this condition produces a recall value similar to GA\_GS\_SA, this is due to how it creates many segments, which also leads to many inaccurate boundaries.

\begin{table}[]
\begin{tabular}{lrrrr}
\hline
% \rowcolor[HTML]{EFEFEF} 
Condition & \multicolumn{1}{l}{WindowDiff} & \multicolumn{1}{l}{Precision} & \multicolumn{1}{l}{Recall} & \multicolumn{1}{l}{F1} \\ \hline
\textbf{GAv3\_GS\_SA} & \textbf{0.4712} & \textbf{0.7446} & 0.7532 & \textbf{0.7211} \\
GAv3\_IS\_SA & 0.5955 & 0.5851 & 0.7529 & 0.6403 \\ \hline
BASELINE & 0.7624 & 0.5308 & \textbf{0.9270} & 0.6460 \\ \hline
\end{tabular}
\caption{Segmentation performance with and without the sentence linking module.}
\Description{This table shows the effect of the sentence linking module on segmentation performance. GAv3_GS_SA, which includes sentence linking (GS), outperforms the variant without it (GAv3_IS_SA) across all metrics—achieving the lowest WindowDiff and highest F1 score. The BASELINE, which lacks both sentence and action linking, shows the highest recall but suffers in precision and overall segmentation quality.}
\label{tab:techEval-sentenceLinking}
\end{table}

\paragraph{\textbf{With or without the sentence assign module}} Finally, there may be cases where sentences are not directly tied to actions (such as ``evaluation'' or ``planning'') and remain unassigned, but they could potentially be included within a decision step. We see that the module responsible for assigning these sentences (Section \ref{system:workflowSegmentation}-Sentence Assigning) improves segmentation performance (\autoref{tab:techEval-sentenceAssigning}). Specifically we observe that, without this module, WindowDiff increases (GAv3\_GS\_SA: 0.4712; GAv3\_GS\_SI: 0.5652) and precision decreases (GAv3\_GS\_SA: 0.7446; GAv3\_GS\_SI: 0.6555). We observe that the pipeline without this module (GAv3\_GS\_SI) simply includes ungrouped sentences as their own separate segments, which increases the total number of segments. While this method does not produce as many segments as the baseline, GAv3\_GS\_SI still creates more segments than the ground truth, resulting in lower precision and higher recall compared to GAv3\_GS\_SA.
\begin{table}[]
\begin{tabular}{lrrrr}
\hline
% \rowcolor[HTML]{EFEFEF} 
Condition & \multicolumn{1}{l}{WindowDiff} & \multicolumn{1}{l}{Precision} & \multicolumn{1}{l}{Recall} & \multicolumn{1}{l}{F1} \\ \hline
\textbf{GAv3\_GS\_SA} & \textbf{0.4712} & \textbf{0.7446} & 0.7532 & \textbf{0.7211} \\
GAv3\_GS\_SI & 0.5652 & 0.6555 & 0.8296 & 0.7066 \\ \hline
BASELINE & 0.7624 & 0.5308 & \textbf{0.9270} & 0.6460 \\ \hline
\end{tabular}
\caption{Segmentation performance with and without the sentence assigning module.}
\Description{This table shows the effect of the sentence assigning module on segmentation performance. GAv3_GS_SA, which includes sentence assigning (SA), outperforms the variant without it (GAv3_GS_SI) in terms of WindowDiff, precision, and F1. Without sentence assigning, the system over-segments, resulting in higher recall but reduced overall accuracy. The BASELINE again yields the highest recall but underperforms in precision and segmentation quality.}
\label{tab:techEval-sentenceAssigning}
\end{table}

\subsection{Performance of Rationale Evaluation Module}

Accurately evaluating whether a user has provided sufficient rationale for a cognitive decision step is crucial to ensure that \sysname{} only asks necessary clarification questions. In this section, we assess how closely our Rationale Evaluation Module’s (Section \ref{system:rationaleEvaluationModule}) judgments align with human judgments.

\subsubsection{Ground-Truth Dataset and Metrics}

\paragraph{Creating the Ground-Truth Dataset}
From the 52 collected recordings, we created a total of 292 cognitive decision steps, with an average of 48.67 decision steps per participant (SD=15.31, MAX=68, MIN=32). For each of these collected steps, two authors evaluated whether the step contained a strong, weak, or empty explanation to create our ground-truth dataset. First, we randomly sampled 20\% of all the decision steps (60 decision steps) and the two authors independently evaluated all steps in this subset---resulting in a Cohen’s Kappa of 0.705, indicating substantial agreement. Subsequently, each author annotated an additional 20\% of the data and creating a total of 180 annotated decision steps---around 65\% of all collected steps. Among these 180 data points, there were 30 strong explanations, 108 weak explanations, and 42 empty explanations.

\paragraph{Metric}
We calculated the average accuracy of the Rationale Evaluation Module by comparing its predicted evaluations of the explanations for each cognitive step against the ground-truth evaluations.

\subsubsection{Results}
The results indicate that our Rationale Evaluation Module aligns with human evaluations for the decision steps with 80.19\% accuracy (SD=1.4\%). We report accuracy for each explanation type: accuracy was 81.11\% accuracy (SD=1.92\%) for ``Strong explanations'', 77.78\% (SD=1.5\%) for ``Weak explanations'', and 85.71\% accuracy (SD=2.38\%) for ``Empty explanations''.
\section{User study}
\label{evaluation}

Our evaluation aims to address the following research questions:
\begin{itemize}
    \item RQ1: How does \sysname{}'s questioning and clarification process during the user's decision steps help in explaining explicit and implicit decisions? 
    \item RQ2: How is the rationale inferred by \sysname{} received, and what impact does providing such inference have on the self-explanation process?
    \item RQ3: How can the decisions captured in the workflow through \sysname{} and the documented decisions be utilized?
\end{itemize}

To answer these research questions, we conducted a within-subjects user study with twelve interface designers, comparing \sysname{} with a baseline---a think-aloud condition without question-based interaction---to evaluate its effectiveness in knowledge capture. We selected this baseline over retrospective methods (e.g., post-hoc interviews) or self-documentation without verbalization to mitigate memory decay and ensure the capture of implicit knowledge that might be lost after the task. While concurrent verbalization imposes some cognitive load, we prioritized the immediate elicitation of decision rationale, aligning with prior findings that in-situ capture outweighs the disruption of manual documentation~\cite{krosnick2021Think}.

We provided the designers with two different web design tasks, and they performed self-explanation-based web design as a form of design livestreaming\footnote{https://www.youtube.com/watch?v=clSHs94hNNc} using both \sysname{} and the baseline.

\subsection{Participants}
We recruited twelve creative professionals (compensated 50,000 KRW/hour) who either majored in design-related fields (e.g., graphic, industrial, or interface design) and who are currently working as interface or UI/UX designers. 
The participants (7 females and 5 males; MEAN\_age=29.3, SD\_age=3.8) had, on average, 5.5 years of experience in UI/UX design (SD=4.1 years, MAX=15 years, MIN=1.4 years)

\subsection{Task Design and Procedure}
The study was conducted in person. Participants were assigned two distinct web interface design tasks in a knowledge-sharing scenario: one involved designing an HCI lab's website, which required making a wide range of design choices, and the other involved designing a basic product page for an e-commerce website, which entailed fewer design choices and adhered to more predefined rules. We showed several examples of knowledge-sharing videos\footnote{https://www.youtube.com/watch?v=Q2IupOJHrzM} before starting the task to provide context for the live-streaming scenario. To minimize potential priming effects—where participants might unconsciously mimic the explanation style of the examples—we explicitly instructed them to explain their design process in their own unique and natural manner, rather than imitating the videos.

To reflect realistic design workflows, participants prepared reference websites ahead of time and referred to them during the design process. Standardized design assets were provided for both tasks. For the lab website task, this included the lab name, introductory text, and project information. For the e-commerce website task, participants received product images, details, and product category information. 

Following a 15-minute introduction, participants worked for 30 minutes in each condition, either with \sysname{} or the baseline system. After each session, participants completed a survey lasting approximately 5 minutes. The order of tasks and conditions was counterbalanced to mitigate ordering effects. Finally, a 30-minute in-depth interview was conducted to gather in-depth feedback. During the in-depth interview process, we reviewed the participant's documented decisions from the workflow together and discussed usage scenarios regarding how these captured decisions could be utilized either by themselves or by others (e.g., novice designers or collaborators).
\label{evaluation:taskDesign}

\subsection{Measures}
As a quantitative metric, we compared the total number of strong, weak, and empty explanations at the level of each captured decision step—evaluating whether each step included a sufficient rationale. After the study, participants’ answers were processed by our rationale evaluation module (Section \ref{system:rationaleEvaluationModule}) to assess whether the clarification process captured missing rationales. Additionally, in the \sysname{} condition, we measured the acceptance rate of system-inferred rationales during the workflow.

For qualitative evaluation, we incorporated NASA-TLX items \cite{hart1988nasa} into the survey to assess task load and added questions to gauge participants’ perceptions of the self-explanation-based workflow. These questions focused on whether the process clarified their decision-making and underlying rationale, improved the articulation of applied knowledge, promoted reflection on decisions, and fostered metacognitive \cite{livingston2003metacognition} insights. Participants also rated the ease of expressing their knowledge, the perceived burden, and the overall effectiveness of capturing both explicit and implicit knowledge (7-Likert scale). To analyze these non-parametric data, we conducted a Wilcoxon signed-rank test. To control for the false discovery rate (FDR) across the 17 survey items, we adjusted the $p$-values using the Benjamini-Hochberg procedure with $q=0.05$.

During the in-depth interviews, participants compared the two conditions in terms of the knowledge capture process and the cognitive burdens experienced. Also, they discussed how sharing their workflow while receiving system-generated questions differed from similar real-world scenarios (such as giving presentations or teaching over Zoom\footnote{https://www.zoom.com/}). In the \sysname{} condition, we asked about the impact, quantity, and quality of the system’s questions. Participants also considered what knowledge or insights might have been missed without questions and evaluated how capturing their workflow in this manner could benefit both themselves and others. Finally, while reviewing the documented decisions within the workflow, we discussed how these captured decisions could be utilized.
\section{Results}

This section describes the findings of our user study, organized around the three research questions. First, we examine how \sysname{}'s questioning and clarification process supports users in articulating both explicit and implicit rationales of their decision steps (RQ1). Second, we analyze how the inferred rationales are received and how they impact the self-explanation process, particularly in aspects of minimizing cognitive burden and enabling deeper reflection (RQ2). Finally, we explore how the captured and documented decisions can be utilized in various usage scenarios, including by the designers themselves, novice designers, collaborators, and design assistants in the form of generative agents (RQ3).

\begin{figure*}[!ht]
  \centering
  \includegraphics[width=1.00\textwidth]{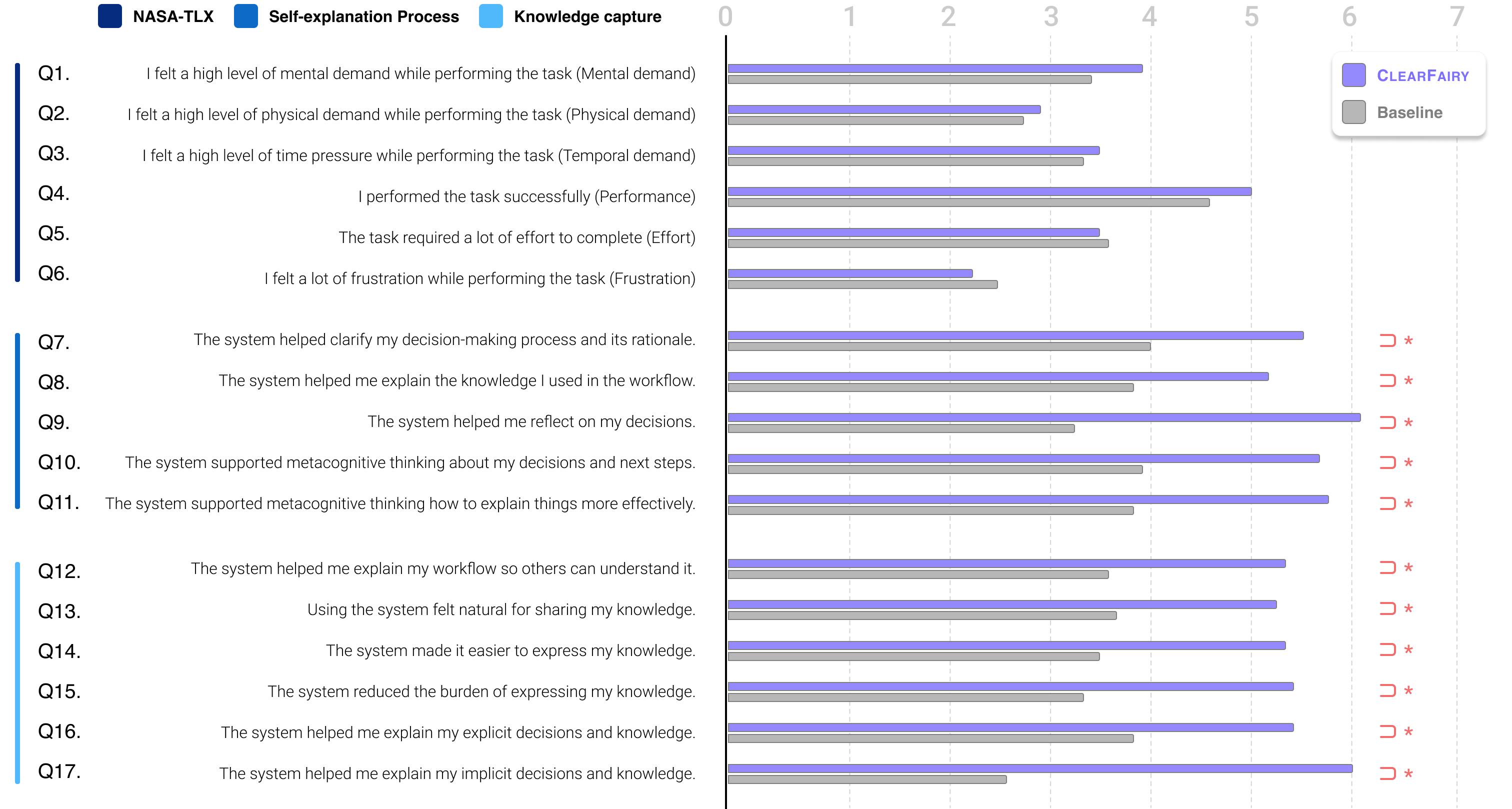}
  \caption{Survey results comparing \sysname{} and the baseline across the metrics related to task load, self-explanation, and knowledge capture. \textasteriskcentered{}: p<0.05}
  \Description{This figure shows participants' ratings on 17 survey questions evaluating system experience across three categories: NASA-TLX workload (Q1–Q6), self-explanation (Q7–Q11), and knowledge capture (Q12–Q17). Bars represent average scores for \sysname{} (purple) and the baseline system (gray). While workload ratings (Q1–Q6) are comparable between conditions, \sysname{} shows significantly higher ratings across all self-explanation and knowledge capture measures.}
  \label{fig:survey}
\end{figure*}

\subsection{RQ1: Capturing Rationales Behind Decisions}
\begin{figure*}[!ht]
  \centering
  \includegraphics[width=1.00\textwidth]{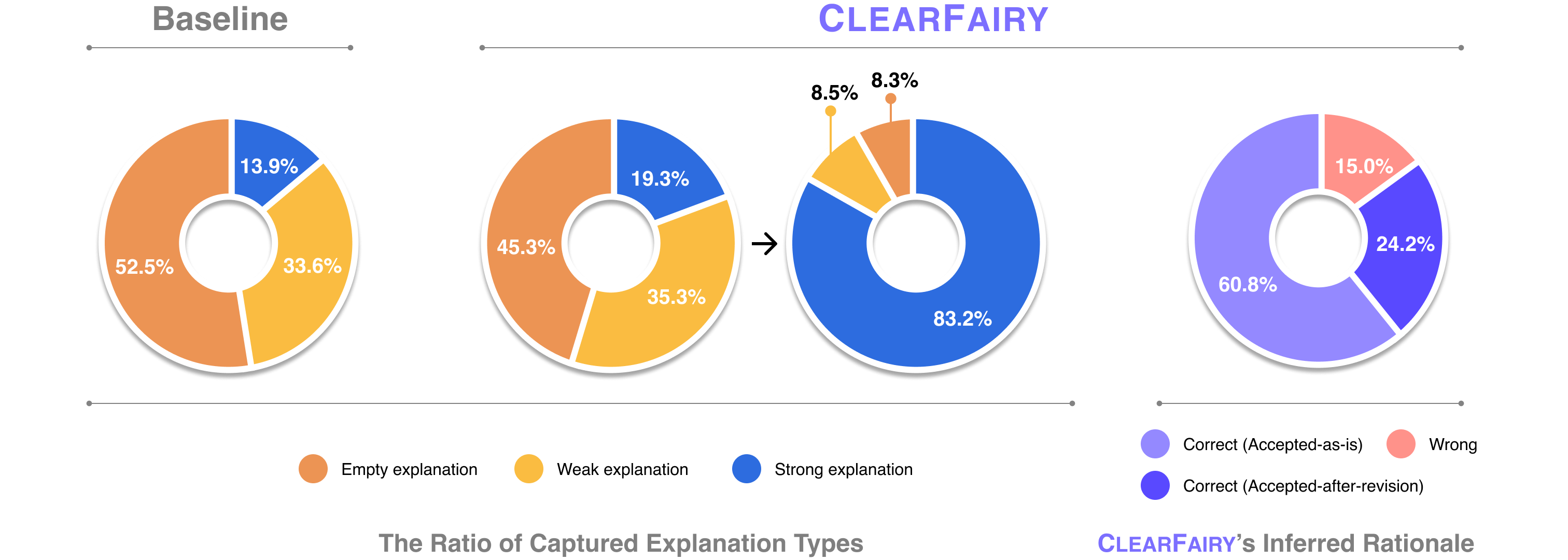}
  \caption{Types of decision steps captured in the Baseline and \sysname{} conditions (Strong, Weak, or Empty), along with user study results on participants’ evaluations of \sysname{}’s inferred rationales. \sysname{} re-evaluates the rationale based on the user’s answer to a clarification question. Participant evaluations of the inferred rationales fall into three categories: Correct (Accepted-as-is), Wrong, and Correct (Accepted-after-revision).}
  \Description{The figure presents four donut charts comparing explanation types in Baseline and ClearFairy conditions and showing participants’ evaluations of inferred rationales. In the Baseline condition, 52.5\% of decision steps were empty explanations, 33.6\% were weak explanations, and 13.9\% were strong explanations. In the ClearFairy condition before inference, 45.3\% were empty, 35.3\% were weak, and 19.3\% were strong. After inference, ClearFairy increased the proportion of strong explanations to 83.2\%, while 8.3\% were weak and 8.5\% were empty. Participants’ evaluations of ClearFairy’s inferred rationales show that 60.8\% were correct, 24.2\% were rationale-eliciting, and 15.0\% were wrong.
}
  \label{fig:quant}
\end{figure*}

Participants in the baseline condition created more decision steps on average (Mean=49.33, SD=14.01) compared to the \sysname{} condition (Mean=34.75, SD=7.44). However, a time-normalized analysis reveals that the rate of step generation was comparable between the two conditions (2.10 vs. 1.80 steps/min), suggesting that the difference in total steps was primarily due to the shorter session duration in the \sysname{} condition. However, through the Q\&A process with participants, \sysname{} transformed an average of 79.13\% (SD=9.52\%) of decision steps that were initially judged as weak or empty. As a result, \sysname{} captured strong explanations containing sufficient rationale for 83.16\% (SD=8.07\%) of all decision steps explained by participants. In contrast, in the baseline condition, only 13.86\% (SD=9.87\%) of the captured decision steps were accompanied by strong explanations (\autoref{fig:quant}-\sysname{}).

Participants reported in the survey that during the self-explanation process, the system’s questions enabled them to explain not only their explicit decisions (\autoref{fig:survey}; Q16, Mean\_diff=1.58, p<0.05) but also those that were implicit decisions that felt so natural to them that they would not have thought to explain them otherwise (\autoref{fig:survey}; Q17, Mean\_diff=3.42, p<0.05). Through this process, participants noted that they were not only able to clarify their rationale more explicitly (\autoref{fig:survey}; Q7, Mean\_diff=1.50, p<0.05), but also articulate the knowledge they were drawing on (\autoref{fig:survey}; Q8, Mean\_diff=1.33, p<0.05). Furthermore, they reported gaining a better understanding of how to provide good explanations (\autoref{fig:survey}; Q12, Mean\_diff=1.75, p<0.05).

Participants (P1, P3, P5–P7, P9, P11–P12) explained that the process enabled them to articulate implicit decisions that had previously felt too, natural, intuitive, or self-evident to describe. Also, some participants (P2, P3, P5–P10) added that the questions corresponded well with their decision units and encouraged them to reflect more deeply on the reasoning behind their decisions. After going through this Q\&A process, they also found it easier to anticipate the kinds of questions that might follow based on the explanations they had provided. 

\sysname{}'s Q\&A process revealed several implicit knowledge types that informed design decisions. For instance, decisions based on \textit{expert conventions} were noted when participants explained that placing key elements—such as title or crucial navigation tabs—in the upper-left corner aligns with standard reading patterns and common industry practices. Similarly, \textit{aesthetic intuition} played a role, with participants describing adjustments to button colors, background hues, or text highlights in order to maintain visual harmony and convey a refined design tone. Some decisions were influenced by \textit{personal preferences}, where designers opted to retain familiar system elements or choose device-specific layouts that best suited their individual tastes. Additionally, \textit{tool usage experience} was evident as participants detailed leveraging features like Figma's auto-layout hug setting and component definition settings in their design software to manage spacing efficiently and adapt to varying content sizes. Decisions regarding \textit{the design process itself} emerged when participants emphasized setting up frames, grids, or transitioning to subsequent design phases as foundational steps.
Other captured knowledge types included \textit{efficiency strategies}, \textit{application of design principles}, and \textit{trial-and-error processes}, such as reusing and adjusting previous components, refining visual details, and comparing layout alternatives.

% \textit{Efficiency considerations based on past experience} were also captured, with designers often replicating or reusing components to expedite their workflow. Moreover, several participants invoked \textit{design theory and principles}—such as adjusting typography and spacing—to enhance readability and establish a clear visual hierarchy. Finally, a \textit{trial-and-error approach} along with \textit{comparative evaluations of alternative options} was employed, as participants experimented with grid configurations, font sizes, and element placements until achieving a visually balanced solution.

\subsection{RQ2: Supporting Effective Knowledge Sharing Without Added Burden}

The Q\&A process with \sysname{} helped participants articulate their rationales more easily and reduced the burden typically associated with knowledge sharing (\autoref{fig:survey}; Q15, Mean\_diff=2.08, p<0.05). Participants (P1, P6-P8, P10-P11) noted that, compared to explaining continuously on their own in the baseline condition, the prompted questions provided moments to pause and reflect more deeply (\autoref{fig:survey}; Q9, Mean\_diff=2.83, p<0.05). Despite feeling there were more questions than typically asked by audiences, all participants found the burden low due to the system’s inference support. There is also no significant difference in NASA-TLX survey results (\autoref{fig:survey}; Q1-Q6, p>0.05).

In the workflow, \sysname{} generated an average of 20.25 inferred rationales (SD=8.59), accounting for 68.44\% of weak and empty explanations. Importantly, participants did not blindly accept the system’s inferences: knowing that their responses would influence the quality of future inferences, they carefully reviewed each one—on average supplementing 24.20\% (SD=15.71\%) and rejecting 15.01\% (SD=6.68\%) (\autoref{fig:quant}-``Accepted-after-revision'' and ``Wrong'' cases). While a few participants (P4 and P10) felt that some inferences were slightly repetitive due to limited initial input, the majority (P1–P3, P5–P9, P11–P12) emphasized that these inferences helped them refine their reasoning. They often treated them as \textbf{rationale-eliciting inferences}, using the system’s plausible suggestions as prompts to elaborate more detailed and nuanced rationales, which represented 24.2\% (SD=15.71\%) of all inference cases (\autoref{fig:quant}-\sysname{}'s Inferred Rationale).

Participants also found the interaction natural (\autoref{fig:survey}; Q13, Mean\_diff=1.58, p<0.05), particularly appreciating the ability to control response timing and resume their workflow smoothly. As they provided richer explanations, subsequent inferences became more specific and aligned with their thinking, creating a productive feedback loop. This loop also enabled participants to surface and articulate hidden rationales that they would not have explained otherwise, without adding extra burden. This process not only reduced the burden of knowledge-sharing but also deepened reflection: participants gained greater awareness of what they knew, the direction they were pursuing, and how they might proceed with their work (\autoref{fig:survey}; Q10, Mean\_diff=1.75, p<0.05; Q11, Mean\_diff=1.92, p<0.05), highlighting the system’s potential to support metacognitive insight~\cite{livingston2003metacognition}.

\subsection{RQ3: Diverse Usage Scenarios of Decisions Captured by \sysname{}}
\label{results:RQ3}
To explore how captured decision steps could be used in different contexts during the interview session, we discussed the potential value of documentation (\autoref{fig:system-documentation}) from multiple perspectives: designers themselves, novices, collaborators, and design agents.

\paragraph{From the Designer's Own Perspective} Participants found the documented decision steps helpful for revisiting their own design rationale, especially during revision or when facing design challenges (P3, P4, and P7, P10-P11). Specifically, participants (P1, P4, and P5) emphasized the value of reviewing their major repetitive decisions or pivotal choices that significantly shifted the project's direction. Some viewed it as a simulation before explaining their work to others (P2 and P12), and others reflected on how to streamline their future workflows (P6). Several participants expressed interest in reusing the captured intuitions for future projects (P3, P8-P9).

\paragraph{From the Perspective of Novice Designers} Participants believed the documentation could help novices learn not just outcomes but also trial-and-error processes and decision-making strategies (P1, P3, P6, and P10). Reflecting on their own novice days, participants (P5, P9, P11-P12) emphasized that observing an expert's micro-decisions is crucial for learning. In particular, they noted that seeing habitual actions accompanied by rationales---such as strategies for efficiency or future maintainability---would allow novices to follow the workflow step-by-step. It was seen as a valuable educational tool to understand how others tackle design problems, although filtering might be necessary due to information density (P2, P7, and P11-P12).

\paragraph{From the Perspective of Collaborators} Participants noted that understanding the reasoning behind design decisions could improve collaboration, especially when building shared design systems or communicating with clients or developers (P1, P5, P7-P9). Documentation of decisions enabled collaborators to trace back choices, reducing unnecessary trial and error (P3-P4, and P10). Regarding the granularity for collaboration, participants (P3, P6-P8, and P12) expressed a preference for viewing key decisions that differed significantly from their own approaches, or high-level summaries of decisions, rather than raw details.

\paragraph{From the Perspective of Design Agents} Participants explained that if agents could access their cognitive decision steps, they could learn personal preferences and offer better reference-based suggestions (P1, P9, and P12). Repetitive tasks like alignment or arrangement layout could be automated in a personalized way (P2, P3, P5, and P12). Others wanted agents to propose multiple alternatives and receive feedback, like a Github Copilot in their process (P3, P4, P6-P7, and P10-P12). Some suggested that by understanding their patterns, agents could also support more creative directions (P8-P9, and P12).

\section{Exploratory Applications of Captured Cognitive Decision Steps}
To demonstrate the potential utility of the detailed cognitive decision steps captured by \sysname{}, this section explores two exploratory applications: predicting a professional's next action (decision and rationale) in a creative workflow, and generating actual artifacts with Model Context Protocol~\footnote{\url{https://modelcontextprotocol.io/introduction}}-based design agents with tool-calling capabilities in Figma. Recent work in agentic data collection highlights the critical need for rich, human-grounded datasets to improve agent reasoning and alignment~\cite{yang2025ultra, bai2025digiq, zhou2024proposer}.

While prioritizing predictability in creativity support tools carries the risk of establishing a constraining ``normative ground''~\cite{li2023beyond}, we utilize agent alignment here specifically as a proxy to validate the system's capacity for \textit{context-aware assistance}. Rather than enforcing a standardized workflow, this evaluation demonstrates that capturing sufficient rationale allows agents to align with and respect the user's \textit{unique intent} and specific workflow—providing tailored support distinct from generic or normative standards.

\subsection{Next Decision Step Generation}
\label{application:nextStep}
The cognitive decision steps captured through \sysname{} provide rich information about a user's unique workflow. This information could be leveraged by generative agents~\cite{huang2024understanding, park2023generativeAgents} to infer the user's next action, enabling context-aware shadowing support—such as proactive suggestions or automated actions similar to GitHub Copilot~\footnote{https://code.visualstudio.com/docs/copilot/overview}—as suggested by our study participants (Section \ref{results:RQ3}).

To investigate whether the captured rationale effectively conveys the user's specific context to the model, we conducted a small-scale exploration comparing predictions made with and without these decision steps. This comparison aims to verify if the rationale helps the agent align with the professional's specific reasoning path rather than defaulting to generic next steps for context-aware shadowing support.

\subsubsection{Approach}
Participants in the user study performed two different web design tasks under two different conditions (Section \ref{evaluation:taskDesign}). From the decision steps collected in the baseline condition, we randomly selected five instances of consecutive pairs of steps.

For each pair, we generated next-step summaries in three ways: (1) without any information (Baseline), (2) using captured decision steps from the \sysname{} condition but excluding the user’s clarified rationale data (Without Rationale), and (3) including the user’s clarified rationale data from the \sysname{} condition (With Rationale). Participants were then shown summaries of the two preceding steps (see the examples in Appendix \ref{appendix:nextStep_example}) along with relevant workflow screenshots and asked to select which of the three steps presented the most suitable ``next step'' in the workflow. Specifically, they were instructed to pick the next step they would most likely take. Each participant completed this selection process for five distinct cases and additionally provided a long-form explanation about why they did not choose the other options.

\subsubsection{Observations}
Across the five survey cases per participant, the baseline was selected an average of 1.00 times (SD = 0.74). Participants who did not choose the Baseline in certain cases remarked that it felt too different from the previous steps or contained ``hallucinated'' decisions or actions. In the cases where the baseline was chosen, participants perceived it included more plausible and high-level decisions (not specific), even if it lacked certain detailed descriptions of actions.

Meanwhile, the ``With Rationale'' condition was chosen 3.33 times on average (SD = 0.65). Participants who chose ``With Rationale'' noted that the ``Without Rationale'' condition’s rationale seemed insufficient or inappropriate for them, whereas those who rejected ``With Rationale'' viewed it as having fewer or more minimal actions, making it appear less comprehensive. The ``Without Rationale'' version was chosen on average 0.67 times (SD = 0.65).

We observed that when the next step was generated based on captured decision steps, it tended to produce finer-grained decisions. Furthermore, in the ``With Rationale'' condition, the decision step appeared to reflect the rationale of the user who had shared it. While generated decisions in the baseline exhibited the highest level of granularity, they tended to yield coarse-grained decisions, sometimes abruptly shifting to a different sub-task without completing the current one. In contrast, when the next steps were generated using the actual decision steps (in both ``With Rationale'' and ``Without Rationale'' conditions), no hallucinations were reported.

\subsection{Design Artifact Creation with Tool-Calling Agent}
We investigate whether incorporating human decision steps accompanied by sufficient rationale into the agent’s design generation process results in artifacts that better reflect the user's intended style and structure. This exploration aims to examine whether providing an agent with professionals’ cognitive decision steps can guide its tool-based design process to produce final artifacts that align with the nuanced ``professional standard'' defined by the user, rather than a generic templates generated by the model's base knowledge.

\label{application:mcp}
\subsubsection{Approach}
We implemented a basic agent that can design directly in Figma. Specifically, we leverage the implementation of the Figma plugin, ``Talk to Figma\footnote{\url{https://github.com/sonnylazuardi/cursor-talk-to-figma-mcp}}'', which provides a Model Context Protocol~\footnote{\url{https://modelcontextprotocol.io/introduction}} that allows agents to call ``tools'' in Figma (e.g., commands or actions like \texttt{create\_rectangle}) through Figma's API (see the detailed propmt in Appendix \ref{appendix:tool-calling_agent_prompt}). Our agent can call 49 specialized tools (Appendix \ref{appendix:tool_list}), which are organized into 6 functional categories (e.g., Style Tools) to enable comprehensive design automation. We implemented the agent based on the ReAct structure~\cite{yao2023react}, which prompts the LLM to perform structured reasoning-action loops with a self-feedback module (Appendix \ref{appendix:feedback_module_ReAct}), guiding the agent to reason about and refine its outputs through multiple iterations. We carefully engineered the prompts to guide the agent by including Figma API usage.

We explored two conditions for design artifact generation: one where the tool-calling agent operated with only the task prompt, and another where we additionally provided design style guidelines extracted from captured decision steps, enabling the agent to align outputs with professionals’ style. To explore this, we conducted artifact generation across tasks at various granularities, including component-level (e.g., subscribe button, navigation menu buttons), section-level (e.g., navigation bar, hero section), and page-level designs (e.g., video streaming main page, startup landing page).

\subsubsection{Observations}
\begin{figure*}[!ht]
  \centering
  \includegraphics[width=1.00\textwidth]{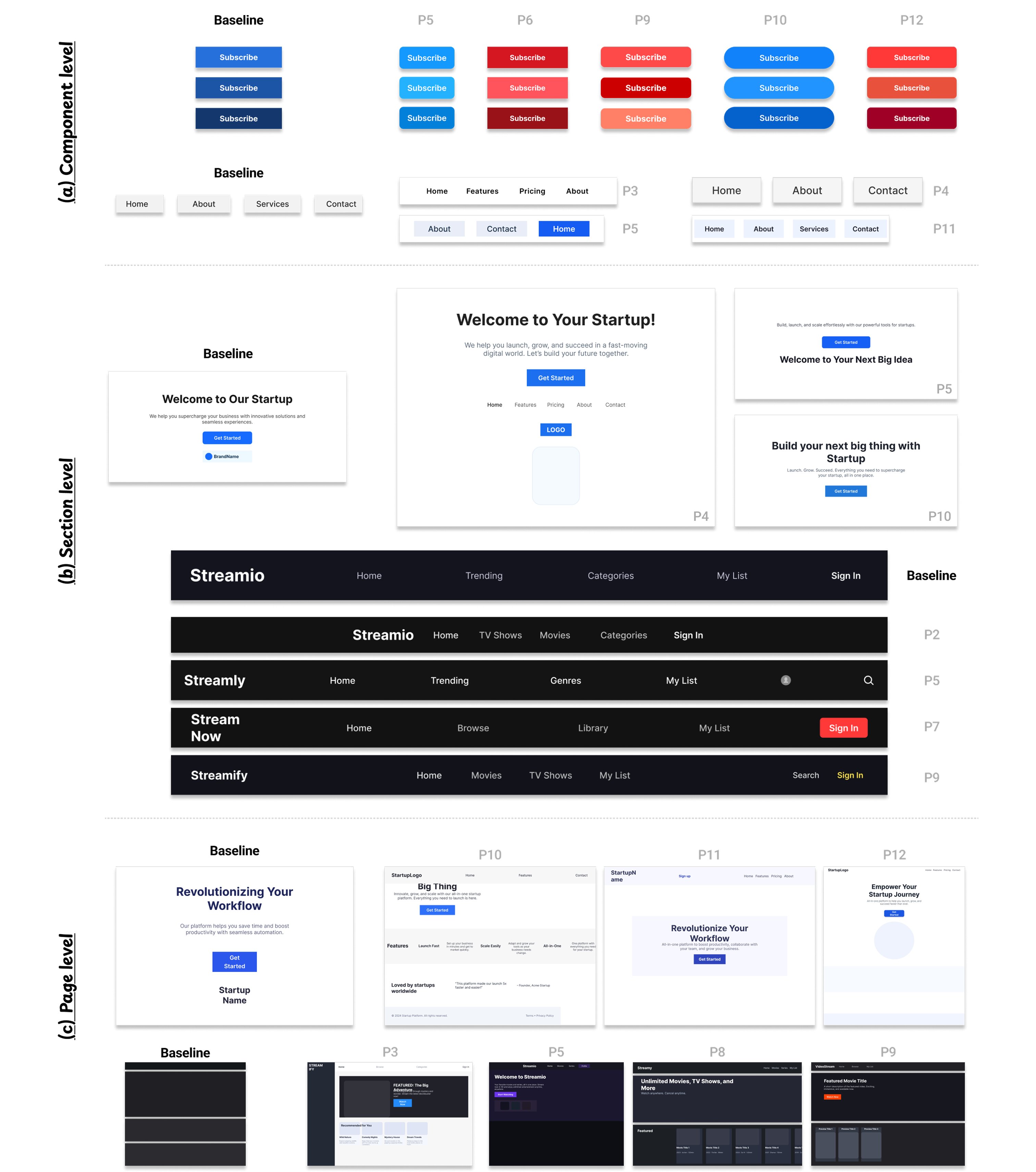}
  \caption{Design outcomes generated with and without the captured Cognitive Decision Steps. The design agent performed six web design tasks across three levels of granularity: component (Subscribe button, navigation menu buttons), section (navigation bar, hero section), and page (video streaming main page, startup landing page).}
  \Description{The figure compares design outcomes generated by the tool-calling agent with and without captured Cognitive Decision Steps across three levels of granularity. At the component level, examples include subscribe buttons and navigation menu buttons; at the section level, examples include a hero section and a navigation bar; and at the page level, examples include a startup landing page and a video streaming main page. Baseline results appear simpler and less structured, while outcomes guided by captured decision steps show more refined layouts, consistent styling, and componentized elements.}
  \label{fig:mcp}
\end{figure*}

\autoref{fig:mcp} shows the overall result of the generated artifacts by our agent. Depending on the given style guidelines from decision steps, the generated artifacts exhibited different styles specifically, regarding the size, color, or layout. An interesting observation is that when a professional’s decision step captured the process of componentizing an element to reuse the element (e.g., button), this appeared to be reflected in the extracted design guidelines, and the agent correspondingly generated the button in a component form (\autoref{fig:mcp}-a, P4). Another example is when the guideline indicated that color should be used to emphasize certain text; we observed that this choice was visibly incorporated into the resulting artifact (\autoref{fig:mcp}-b, P9).

Overall, the agent occasionally produced incomplete layouts, particularly at higher levels of granularity where more complex artifacts were required (\autoref{fig:mcp}-c). However, outputs guided by decision step–derived guidelines tended to incorporate noticeably more core elements compared to the baseline without such guidelines (\autoref{fig:mcp}-c).

In this exploration, we observed that even subtle but distinct variations in style and content emerged in the generated artifacts depending on the captured decision steps. At the same time, we also found that the tool-calling agent’s overall performance in designing on the canvas (e.g., using frame auto-layout much like a human) was limited. Its effectiveness appeared contingent upon how the agent itself was designed, including factors such as prompt engineering, the range and type of available tools, and the inclusion of feedback modules within the agent pipeline.

\section{Discussion}
This section describes the broader implications of capturing more implicit human cognitive decision steps, focusing on their potential applications in supporting learning and advancing generative agents, as well as the generalizability and limitations of our approach.

\subsection{Beyond Natural Language Self-Explanation: From Verbalization to Multimodal and Action-Based Inference}

While many participants found \sysname{} helpful for articulating implicit reasoning, participants (P1, P4, P6, and P11) noted that certain decisions were difficult to explain using natural language alone. Instead, they expressed a preference for referencing previously recorded decision steps or directly annotating specific areas of the current artifact (similar to Stylette's approach~\cite{kim2022stylette}) to convey their thinking and perspective. 
This reflects the nature of creative workflows, where many decisions are deeply embedded in the contextual fabric of the work~\cite{fischer2020making}. To better support such workflows, future work can consider alternative response modalities beyond language---such as direct visual annotations, referencing past interactions, or selecting from visual examples.

Furthermore, our current reliance on concurrent verbalization highlights a critical trade-off in knowledge capture baselines. While we prioritized in-situ verbalization to ensure high-fidelity rationale capture and mitigate memory decay~\cite{krosnick2021Think}, the ultimate goal is to minimize the user's burden. Currently, verbal explanations serve as essential ground truth to map professional behaviors to their cognitive reasoning in a workflow. As \sysname{} accumulates a user's cognitive decision steps, future iterations could transition towards action-based inference---where the system infers the user's rationale solely from behavioral logs without requiring explicit speech (similar approach with GUM framework~\cite{shaikh2025creating}). In this vision, the system would act as a silent observer, prompting the user only when the inferred rationale is ambiguous, thereby combining the low cognitive load of post-hoc methods with the precision of in-situ capture.

In addition to reducing burden, participants (P2, P3, and P9) suggested going beyond simple rationale clarifications and envisioned \sysname{} conducting a design evaluation-based review of the user’s design, followed by Q\&A to elicit otherwise implicit intentions. While \sysname{} already surfaces implicit decisions through its clarification and inference process, this feedback points to future directions where generative models could extend such capabilities by engaging in UI design review–based~\cite{duan2024generating} questioning to capture hidden layers of reasoning.

\subsection{Applications of Human Cognitive Decision Steps in Learning}
Captured cognitive decision steps---especially those of professionals---present a valuable opportunity to support novices. This aligns with the notion of over-the-shoulder learning, where an expert looks over the shoulder of a novice's workflow, helping them as they accomplish a task~\cite{twidale2005over}. The cognitive apprenticeship framework~\cite{collins1991cognitive, li2024tutorly} also emphasizes modeling expert thinking to support novice understanding. If professional decision steps and their accompanying rationales can be captured as structured data, they could serve as reusable, interactive guidance during novice workflows. Several participants (P1, P3–P7, and P11-P12) mentioned this type of application during interviews, noting its potential to help novices directly observe how professionals navigate a sequence of decisions.
For instance, when a novice encounters a decision point, the system could present similar scenarios in which professionals made cognitive decisions, along with their reasoning. This form of interaction may promote deeper internalization and embodiment of professional strategies and decision-making processes~\cite{son2024demystifying}.

\subsection{Capturing Human Cognitive Decision Steps for Agent Planning and Action}
Human workflows differ fundamentally from agent workflows in how cognitive decisions are formed and translated into action \cite{zora2025howdoai}. Generative agents typically consist of a planning module for generating task sequences, a memory module for referencing prior experiences, and an action module for selecting the next move~\cite{huang2024understanding, park2023generativeAgents}. While recent work aims to strengthen the reasoning process of these agents and utilize it as a foundation for subsequent actions~\cite{sun2024emma, yang2025magma}, the reasoning generated by generative models often remains high-level and lacks the fine-grained detail required for precise decision-making. As demonstrated in our Application section (Section~\ref{application:nextStep}), the actions generated by LLMs are frequently coarse, occasionally skipping intermediate steps. For instance, after creating a button in the top navigation bar, the agent might skip the necessary adjustment of its alignment and prematurely proceed to populate the body content.

In contrast, the cognitive decision steps captured through our system are more fine-grained, reflecting the sequence of micro-decisions that occur during professionals’ workflows. As prior work (e.g., Auto-Intent~\cite{kim2024autoIntent}) demonstrated that LLM-generated intents improved agent performance, we argue that capturing human cognitive decision steps and incorporating them into agents' action or planning modules can similarly enhance the agent’s ability to execute more deliberate, precise, and context-sensitive actions. As shown in our second exploration (Section~\ref{application:mcp}), these finer-grained actions led to more coherent artifacts that aligned more closely with professional outputs.

As demonstrated in our second exploration (Section~\ref{application:mcp}), these finer-grained actions not only improved the agent’s process but also led to final artifacts that were more structured, stylistically consistent, and better aligned with professional outputs.

\subsection{Generalizability of \approachname{} Beyond Creative Workflows}
While \approachname{} was designed with creative workflows for knowledge-sharing and demonstrated through UI/UX design scenarios, the underlying principles are rooted in general cognitive theories. Thus, its applicability extends beyond creative tasks. Any domain that involves iterative decision-making—such as literature reviews in research, idea exploration for innovation, or even collaborative workflows with LLMs—could benefit from structuring and capturing cognitive decision steps.

Technically, extending \approachname{} to these domains requires an environment capable of capturing three synchronized signals: user verbalizations, observable artifact states, and fine-grained action logs (e.g., specific image adjustments like ``Apply Gaussian Blur'' in raster graphics editors, or encoding specifications like ``Map Data to Color'' in visualization tools). Provided that a system supports logging these data streams, \approachname{}'s mechanism of dynamically linking actions and outcomes to rationales can be directly applied to structure decision-making processes in these fields.

Moreover, \approachname{} does not impose a fixed granularity for decision segmentation. Instead, it allows steps to be flexibly structured according to the nature of the task, the user’s workflow, and the user's thought process. This adaptability suggests that \approachname{} has the potential for wide generalizability across domains where fine-grained reasoning and decision tracking are critical.

\subsection{Limitations and Future Work}
This work has several limitations: First, some participants (P4 and P5) noted that the types of questions asked by \sysname{} were often similar. Since the current Question Generation Module focuses solely on rationale clarification, future work should explore a wider variety of question types—such as Socratic Questions~\cite{padesky1993socratic}—to capture cognitive decision steps more comprehensively.

Second, the system is primarily designed to evaluate rationales revealed through self-explanations. However, it is also important to identify and clarify cognitive decision steps based on actions alone, even when no accompanying explanation is provided. Actions can implicitly reflect reasoning, and future work should explore how unlinked actions may convey intent and how such reasoning can be captured.

Third, while \sysname{} is currently implemented as a Figma plugin, it should be extended to support more general creative workflows. Many creative tasks unfold across different software platforms, and applying \approachname{} in such contexts may require developing an OS-level system for broader workflow capture. Expanding beyond the current implementation, we see strong parallels between our approach and the \textit{computational notebook paradigm}~\cite{kluyver2016jupyter, rule2018exploration}. Future work could explore presenting captured decision steps as executable ``cells,'' allowing users to modify parameters and replay workflows. This would effectively bring the benefits of \textit{literate programming}~\cite{knuth1984literate} to creative domains, fostering a deeper culture of knowledge reuse.

Lastly, regarding the concern that proactive questioning or inference might encourage over-reliance, our study observed that participants maintained agency by actively reviewing, supplementing, and sometimes rejecting the system's suggestions rather than accepting them passively. However, as knowledge-capture systems become more automated, the risk of users over-relying on imperfect AI inferences increases. Future work should explicitly address this by incorporating mechanisms to calibrate user trust, such as displaying confidence indicators or designing interaction flows that require user verification for low-confidence inferences.
\section{Conclusion}
This work presented \approachname{} and \sysname{}, which structure creative workflows into cognitive decision steps by linking explanations, actions, and artifacts. In a study with 12 professionals, our system significantly increased strong explanations and surfaced implicit decisions while minimizing burden. We further demonstrated that captured steps can enhance the alignment and quality of generative AI outputs. Together with a released dataset of 417 steps, our work highlights the potential of cognitive decision steps for knowledge sharing and human-aligned AI agents.

% This work introduced \approachname{}, a workflow segmentation approach for structuring creative professionals’ decision-making into cognitive decision steps, and \sysname{}, a think-aloud AI assistant that operationalizes \approachname{} within a creative workflow. By linking self-explanations, actions, and artifacts in real time, our system detects decisions that have weak rationales, generates clarifying questions, and infers missing reasoning to capture richer explanations while reducing the knowledge-sharing burden. In a study with 12 creative professionals, \sysname{} significantly increased the proportion of strong explanations and surfaced implicit decisions while minimizing cognitive burden. Beyond empirical validation, this work explored how captured decision steps can be reused for generative applications, including next-action inference and tool-based design simulation, demonstrating their potential to enhance the alignment, coherence, and professional quality of outputs. We also released a dataset of 417 decision steps to support future research. Taken together, our contributions highlight the potential of cognitive decision steps as a foundation for knowledge sharing, reflective practice, and more deliberate and human-aligned AI agents in creative workflows.

\begin{acks}
This work was funded by the KAIST-NAVER hypercreative AI center
This work was supported by Institute of Information \& Communications Technology Planning \& Evaluation (IITP) grant funded by the Korea government (MSIT) (No.2019-0-00075, Artificial Intelligence Graduate School Program(KAIST)). We thank all of our participants for engaging positively in our various studies. We also thank all of the members of KIXLAB for their helpful discussions and constructive feedback.   
\end{acks}

\bibliographystyle{ACM-Reference-Format}
\bibliography{main}

\appendix
\clearpage
\section{Technical Detail}
\subsection{\approachname{}-based Workflow Segmentation Pipeline}
\subsubsection{Prompt: Sentence Linking}
\label{appendix:prompt_sentence_linking}
\mbox{}\par
\begin{prompt}
\obeylines
\setlength{\parindent}{0pt}
\textbf{Prompt Instruction:}

You are a helpful and precise assistant tasked with segmenting a interface designer's transcript into sets of closely related sentences. The transcript is divided in sentence level. The designer conducted a think-aloud session while designing a web interface in Figma, explaining their thought process, decisions, and adjustments. You are given this transcript data. 

Follow these rules for segmentation. 
1. Carefully review the given transcript data (`transcript`) and understand the design context, considering the goal of the session, tools or functions being discussed, and specific design decisions mentioned. If the transcript contains errors or inaccuracies due to automatic speech recognition (STT) issues, infer the intended meaning based on the overall context to ensure an accurate understanding of the web designer's thought process.

2. Group sentences into a segment when they describe the same design element (e.g., font, layout, frame, button, rectangle) or properties (e.g., size, color, shape), design decision (e.g., I need to edit this element), or a specific design sub-task with a goal (e.g., This is too big, so it would be better to edit this). Each segment should be tightly connected by a clear purpose or task. If a sentence does not explicitly mention the design element being modified or its properties, infer it from the surrounding context in the transcript. Consider prior sentences and the overall design flow to determine what element or property is being referred to, ensuring accurate grouping.

3. Keep segments concise and focused. Aim for grouping multiple sentences closely related to a single design aspect (e.g., desgin action or sub goal). Don't group too many sentences in a group. 

Additional rules:
- DO NOT change the original transcript content when grouping. 
- DO NOT change the sentence order.
- DO ONLY group consecutive sentences.

\textbf{Few shots}
\begin{verbatim}
{{Few shots}}
\end{verbatim}

\textbf{Input:}
\begin{verbatim}
transcript: {transcript in sentence level}
\end{verbatim}

\textbf{Formatting Instruction:}
\begin{verbatim}
{
  "0": "All combined related sentences for group 0.",
  "1": "All combined related sentences for group 1.",
  ...,
  "...": "..."
}
\end{verbatim}
\end{prompt}

\subsubsection{Prompt: Sentence and Action-Artifact Linking}
\label{appendix:prompt_sentence_action-artifact}
\mbox{}\par
\begin{prompt}
\obeylines
\setlength{\parindent}{0pt}
\textbf{Prompt Instruction:}

You are a helpful and precise assistant tasked with linking an interface designer's segmented workflow transcripts (\texttt{segmented\_transcripts}) to their corresponding action logs (\texttt{sets\_of\_design\_action\_and\_screenshot}) in Figma. These logs are accompanied by workflow screenshots \\(\texttt{sets\_of\_design\_action\_and\_screenshot}). In Figma, the interface designer's actions (e.g., moving or adjusting design elements such as text or rectangles) are recorded as logs. You will be provided with Figma log data recorded at each timestamp. Each log is paired with a screenshot captured at the same timestamp, offering visual context to verify and validate the linkage between the transcripts and logs.

Your task is to establish links between the segmented transcripts \\(\texttt{segmented\_transcripts}) and the action log's timestamps \\(\texttt{sets\_of\_design\_action\_and\_screenshot}). Follow these steps and rules:
1. Carefully review the given segmented transcripts (\texttt{segmented\_transcripts}) to understand the designer's workflow.
2. Examine the given action log data (\texttt{sets\_of\_design\_action\_and\_screenshot}) to understand the design workflow context.
3. Link each transcript segment to the relevant timestamp(s) of action log(s) when the transcript describes or relates to the action or design context captured in the summarized log and screenshot.
4. A single transcript segment may be linked to multiple timestamp(s) of action log(s) with screenshot, and conversely, a single log with a screenshot may connect to multiple transcript segments.
5. Do not simply link based solely on the element names mentioned in the logs; instead, review the corresponding screenshots and logs together, and consider the preceding and following context (action logs and screenshots) in the transcript. Only link when there is a clear and genuine association between the transcript segment and the log data.

\textbf{Input:}
\begin{verbatim}
segmented_transcripts: {segmented_transcripts}
sets_of_design_action_and_screenshot: {sets_of_design_action_and_screenshot}
\end{verbatim}

\textbf{Formatting Instruction}:
\begin{verbatim}
{
    "links": [
        {
            "<timestamp>": [
                "<related_sentence_index>",
                "…",
                "<related_sentence_index>"
            ]
        },
        "…",
        {
            "<timestamp>": [
                "<related_sentence_index>",
                "…",
                "<related_sentence_index>"
            ]
        }
    ],
    "reversed_links": [
        {
            "<related_sentence_index>": [
                "<timestamp>",
                "…",
                "<timestamp>"
            ]
        },
        "…",
        {
            "<related_sentence_index>": [
                "<timestamp>",
                "…",
                "<timestamp>"
            ]
        }
    ]
}
\end{verbatim}
\end{prompt}

\subsubsection{Prompt: Sentence Assigning}
\label{appendix:setence_assigning}
\mbox{}\par
\begin{prompt}
\obeylines
\setlength{\parindent}{0pt}
\textbf{Prompt Instruction:}

You are a precise assistant tasked with determining the appropriate direction for an unassigned sentence in a web designer's transcript. You will be provided with:
1. The full transcript (transcript) to understand the overall context.
2. An unassigned grouped sentence (unassigned\_grouped\_sentence) that needs classification.
3. The adjacent grouped sentences: left\_grouped\_sentence (immediately preceding) and right\_grouped\_sentence (immediately following).

Your task:
- Analyze the full transcript to grasp the designer's workflow and context.
- Evaluate the semantic relationship of unassigned\_grouped\_sentence with the left\_grouped\_sentence and right\_grouped\_sentence.
\quad- If the unassigned\_grouped\_sentence is more semantically aligned with the left\_grouped\_sentence, output \"left\".
\quad- If it is more aligned with the right\_grouped\_sentence, output \"right\".
\quad- If it does not clearly relate to either, output \"unrelated\".
    
Additional guidelines:
- If the unassigned sentence describes an evaluation or the completion of an action or design element, lean towards 'left'.
- If it describes a plan or intention regarding an action or design element, lean towards 'right'.
- If either left\_grouped\_sentence or right\_grouped\_sentence is empty, base your decision solely on the available non-empty context.
- If both adjacent contexts are empty, output 'unrelated'.
- Base your decision on the overall flow and semantic context of the transcript. 

- If the decision made in the unassigned sentence contradicts the decision in the left or right context, classify it as 'unrelated'.
\quad- A contradiction occurs when the unassigned sentence refers to a different design element or a different property of an element than the left or right context.
\quad- Even if the same element or property is discussed continuously, if the purpose or outcome is modified in the unassigned sentence, classify it as 'unrelated'.
- If the decision made in the unassigned sentence is a continuous extension of the left or right context, classify it as 'left' or 'right' accordingly.
\quad- A continuation occurs when the unassigned sentence builds upon or directly follows the decisions made about the same element or property in the adjacent context.

Please answer 'left', 'right', or 'unrelated'.

\textbf{Input:}
\begin{verbatim}
transcript: {transcript}
unassigned_grouped_sentence: {unassigned_grouped_sentence}
left_grouped_sentence: {left_grouped_sentence}
right_grouped_sentence: {right_grouped_sentence}
\end{verbatim}
\end{prompt}

\subsection{Rationale Evaluation Module}
\subsubsection{Prompt: Rationale Evaluation}
\label{appendix:prompt_rationale_evaluation}
\mbox{}\par
\begin{prompt}
\obeylines
\setlength{\parindent}{0pt}
\textbf{Prompt Instruction:}
You are a precise and analytical assistant tasked with analyzing the workflow of a web designer and assessing their think-aloud explanations of their design process. Specifically, you will be provided with an explanation that a designer provided for a decision step in their process. You should categorize the explanation into various types depending on its content and depth. You will review a web designer's web design think-aloud transcript in Figma.

Instruction.
For the designer's explanation, you should determine which type of explanation it falls under. Below, you will be presented with a list of definitions for the different types of explanations.

\begin{itemize}
  \item \textbf{Strong Situation Relevance (S-SR)}: The explanation connects the decision to a specific task, design goal, or situational context. Focus on whether the reasoning explains the broader purpose of the action beyond just the immediate decision.
  
  \item \textbf{Strong Prior Knowledge-Based (S-PK)}: The decision is justified using established principles, theories, expertise, or experience that are clearly contextualized. Personal preference is only valid if explained in relation to the design context.
  
  \item \textbf{Strong Consideration of Alternatives (S-CA)}: The designer compares multiple options and explicitly explains the reasoning behind their final choice. It must include a decision factor or rationale for choosing one alternative over another.
  
  \item \textbf{Weak Situation Relevance (W-SR)}: The reasoning is too obvious, 
  self-evident, or lacks depth. W-SR includes surface-level causality (e.g., ``because it's too big, I'll make it smaller'') without deeper explanation.
  
  \item \textbf{Weak Prior Knowledge-Based (W-PK)}: References experience, heuristics, or preference without explaining why they are relevant to the current design. Personal preferences are weak unless justified within the design context. Explanations that include personal evaluations related to aesthetic or subjective feelings, such as ``I like/don't like it'', ``Looks good/ugly'' are considered W-PK if no clear reasoning is provided about why these evaluations are relevant to the current situation.
  
  \item \textbf{Weak Consideration of Alternatives (W-CA)}: Mentions alternative choices but fails to explain why one option is preferred. Simply stating that a decision is being made without comparison dimensions is weak. If a designer says ``I'll try this and see,'' but does not finalize a decision, classify as W-CA (because they acknowledge alternatives but don't justify a choice).
  
  \item \textbf{Empty (E)}: An empty explanation does not attempt to justify the decision at all. It simply describes the action or thought process without providing reasoning. A key distinction with weak explanations is that weak explanations at least attempt to provide a reason (even if superficial), whereas empty explanations offer no reasoning whatsoever.
\end{itemize}

Your Task

You will be provided with a segment of a transcript in which a web designer explains a decision. You should categorize it into the types above by following these steps:
\begin{enumerate}
  \item Familiarize yourself with the explanation and analyze its content and depth.
  \item Compare the explanation against each of the definitions for the types provided above. Explain your reasoning as you consider whether the explanation falls under each category.
  \item Decide on the appropriate categories for the explanation based on the definitions provided. Select \textbf{all} categories that fit the overall content and depth of the explanation. You can select multiple categories. You must assign \textbf{at least one} category. When you assign \textbf{Empty}, you should include ``E'' in the output categories.
\end{enumerate}

\textbf{Few shots}
\begin{verbatim}
{{Few shots}}
\end{verbatim}

\textbf{Input:}
\begin{verbatim}
decision_step: {decision_step}
\end{verbatim}

\textbf{Formatting Instruction}:
\begin{verbatim}
{
  "reason": "<Reason for the judgment>",
  "categories": ["category1", "category2", ...]
}
\end{verbatim}
\end{prompt}
\subsection{Question Generation and Rationale Inference Module}
\subsubsection{Prompt: Question Generation}
\label{appendix:prompt_question_generation}
\mbox{}\par
\begin{prompt}
\obeylines
\setlength{\parindent}{0pt}
\textbf{Prompt Instruction:}
You are a helpful and analytical assistant tasked with asking questions to clarify a web designer's design rationale. A design rationale is an explanation of how and why an artifact—or a specific part of it—is designed the way it is. It offers more than just describing the decision; it reveals the underlying reasons, justification, alternatives considered, and the argumentation that led to the decision. In short, a design rationale explains \textbf{why} a design exists in its current form by describing what the artifact is, what it is supposed to do, and how it evolved to this point.

You will be provided with a web designer's think-aloud explanation on their design workflow (\texttt{Explanation}). Also you will be provided a decision step (this is a part of the explanation that contains decisions) along with:
\begin{itemize}
  \item The type of explanation (\texttt{weak} or \texttt{empty}) for the decision step
  \item The reason why this decision step lacks sufficient rationale
\end{itemize}

Instructions
\begin{enumerate}
  \item \textbf{Deeply analyze} the given explanation, the explanation type, and the reason why it lacks rationale. Consider the design workflow from which this explanation likely came.

  \item \textbf{Review the definitions of explanation types carefully}, paying attention to the criteria for strong, weak, and empty explanations.
\end{enumerate}

\textbf{Definitions}
Strong Explanation
Definition: A strong explanation prioritizes the reasoning behind a decision over just describing the outcome. It explicitly provides context, justification, or rationale tied to the design task, principles, or audience. A strong explanation falls into one of the following categories:
\begin{itemize}
  \item \textbf{Strong Situation Relevance (S-SR)}: The explanation connects the decision to a specific task, design goal, or situational context. Focus on whether the reasoning explains the broader purpose of the action beyond just the immediate decision.

  \item \textbf{Strong Prior Knowledge-Based (S-PK)}: The decision is justified using established principles, theories, expertise, or experience that are clearly contextualized. Personal preference is only valid if explained in relation to the design context.

  \item \textbf{Strong Consideration of Alternatives (S-CA)}: The designer compares multiple options and explicitly explains the reasoning behind their final choice. It must include a decision factor or rationale for choosing one alternative over another.
\end{itemize}
Weak Explanation
Definition: A weak explanation provides some reasoning, but it is either superficial, lacks depth, or fails to justify why the choice is appropriate. Weak explanations fall into one of the following categories:
\begin{itemize}
  \item \textbf{Weak Situation Relevance (W-SR)}: The reasoning is too obvious, 
  self-evident, or lacks depth. W-SR includes surface-level causality (e.g., ``because it's too big, I'll make it smaller'') without deeper explanation.

  \item \textbf{Weak Prior Knowledge-Based (W-PK)}: References experience, heuristics, or preference without explaining why they are relevant to the current design. Personal preferences are weak unless justified within the design context. Explanations that include personal evaluations related to aesthetic or subjective feelings, such as ``I like/don't like it'', ``Looks good/ugly'' are considered W-PK if no clear reasoning is provided about why these evaluations are relevant to the current situation.

  \item \textbf{Weak Consideration of Alternatives (W-CA)}: Mentions alternative choices but fails to explain why one option is preferred. Simply stating that a decision is being made without comparison dimensions is weak. If a designer says ``I'll try this and see,'' but does not finalize a decision, classify as W-CA (because they acknowledge alternatives but don't justify a choice).
\end{itemize}
Empty Explanation
Definition: An empty explanation does not attempt to justify the decision at all. It simply describes the action or thought process without providing reasoning. A key distinction with weak explanations is that weak explanations at least attempt to provide a reason (even if superficial), whereas empty explanations offer no reasoning whatsoever.

\begin{enumerate}
  \setcounter{enumi}{2}
  \item Based on the provided explanation and its type, \textbf{identify which weak or empty categories it falls into} by referring to the definitions above.

  \item \textbf{Generate a short, clear, and good rationale clarification question} (that considers the design context) designed to elicit a deeper, more thoughtful explanation from the designer. The question should encourage the designer to reflect on their design rationale in a way that could transform the current weak or empty explanation in the \texttt{decision\_step} into a strong explanation.
\end{enumerate}

\textbf{Input:}
\begin{verbatim}
explanation: {explanation}
decision_step: {decision_step}
reason_for_insufficient_rationale: {reason_for_insufficient_rationale}
\end{verbatim}

\textbf{Formatting Instruction}:
\begin{verbatim}
{
  "question": "<generated question>"
}
\end{verbatim}
\end{prompt}

\subsubsection{Prompt: Rationale Inference}
\label{appendix:prompt_rationale_inference}
\mbox{}\par
\begin{prompt}
\obeylines
\setlength{\parindent}{0pt}
\textbf{Prompt Instruction:}
You are an analytical and precise assistant tasked with inferring a web designer's rationale behind their decision based on their previous decisions with rationales. The designer conducted a think-aloud session while designing a web interface in Figma, explaining their thought process, decisions, and adjustments. You will be provided with 1) the full transcript (`Full transcript`) of the session (for overall session context) that includes the decision\_step's transcript, 2) a specific segment of the transcript containing the designer's decision steps (`Decision step`), 3) the reason (`Reason for insufficient rationale`) why the decision\_step in 2) lacks a rationale or has an insufficient rationale, and 4) information on the designer's previous decisions and each decisions rationale (`Previous decision step summaries`).

Your Task
You must infer the rationale of the given decision\_step through the following steps:

Step 1: First, understand the design context based on the provided full transcript (`Full transcript`) and carefully review the decision step (`Decision step`).
Step 2: Carefully review the information about the reason why this decision\_step lacks sufficient rationale (`Reason for insufficient rationale`).
Step 3: Then, review the provided previous design decisions and their rationale information (`Previous decision step summaries`). Each summary contains the following: `decision\_and\_actions` (A single-line sentence summarizing the core design decisions and actions taken by the designer), `rationale` (A single-line sentence explaining the reason behind the designer's key decisions), and `progression` (A single-line sentence describing how the design process logically progressed based on the designer's decisions).
Step 4: Review each summary one by one, and select the summaries that have `decision\_and\_actions` or `progression` directly similar to the given decision\_step. Summaries should be considered similar only if the specific situations, elements being modified, or actions taken are literally the same or equivalent. You must not consider summaries similar based on abstract or high-level reasons, such as both aiming to improve user experience, having similar design goals, or sharing general intentions. Abstract similarity is not sufficient; there must be a concrete, direct match in the specific decisions and actions taken or the `progression` described.
Step 5: If there are similar summaries, infer the rationale of the given decision\_step based on the `rationale`(s) of the selected summary or summaries.
\textbf{Input:}
\begin{verbatim}
explanation: {explanation}
decision_step: {decision_step}
reason_for_insufficient_rationale: {reason_for_insufficient_rationale}
previous_decision_step_summaries: {previous_decision_step_summaries}
\end{verbatim}

\textbf{Formatting Instruction}:
\begin{verbatim}
{
    "inferred_rationale": "<inferred_rationale or None>",
    "reasoning": "<reasoning of your inference or None>"
}
\end{verbatim}
\end{prompt}

\subsection{Decision Step Summary Module}
\subsubsection{Prompt: Decision Step Summary Generation}
\label{appendix:prompt_decision_step_summary_generation}
\mbox{}\par
\begin{prompt}
\obeylines
\setlength{\parindent}{0pt}
\textbf{Prompt Instruction:}
You are a helpful and precise assistant tasked with summarizing a web designer's decision steps in a think-aloud session. The designer conducted a think-aloud session while designing a web interface in Figma, explaining their thought process, decisions, and adjustments. 

You will be provided with:
\begin{itemize}
  \item The full transcript (\texttt{Full transcript}) of the session (for overall session context) that includes the \texttt{decision\_step}'s transcript,
  \item A specific segment of the transcript containing the designer's decision steps (\texttt{Decision step}),
  \item Related task workflow screenshots (\texttt{Screenshots related to the decision step}) providing visual context for the decision step.
\end{itemize}

Your Task
First, review the given \texttt{Decision step} and full transcript to understand the designer's workflow and their decision. Then, summarize the designer's key decision step into three clear, single-line sentences, each capturing the following:

\begin{enumerate}
  \item \textbf{Main Design Decisions and Actions} (\texttt{decision\_and\_actions} in the output format):  
  A single-line sentence summarizing the core design decisions and actions taken by the designer.
  
  \item \textbf{Rationale} (\texttt{rationale} in the output format):  
  A single-line sentence explaining the reason behind the designer's key decisions.
  
  \item \textbf{Progression} (\texttt{progression} in the output format):  
  A single-line sentence describing how the design process logically progressed based on the designer's decisions.
\end{enumerate}

Guidelines
\begin{itemize}
  \item Focus on the core design decisions and actions taken by the designer with the explained rationale.
  \item Include key details that are essential to understanding what the designer did and why — such as specific properties adjusted (e.g., font size, spacing, layout), components added or removed, or designer's detailed reasoning process.
  \item Do not include extraneous details or unrelated information in the summary.
  \item Use clear and precise language to convey the main points of the designer's workflow.
  \item Ensure that the summary reflects the logical flow and sequence of the decision steps.
\end{itemize}

\textbf{Input:}
\begin{verbatim}
transcript: {transcript}
decision_step: {decision_step}
related_screenshots: {related_screenshots}
\end{verbatim}

\textbf{Formatting Instruction}:
\begin{verbatim}
{
    "decision_and_actions": "<decision_and_actions>",
    "rationale": "<rationale>",
    "progression": "<progression>"
}
\end{verbatim}
\end{prompt}

\subsubsection{Prompt: Decision Step Summary Generation with Answer}
\label{appendix:prompt_decision_step_summary_generation_with_answer}
\mbox{}\par
\begin{prompt}
\obeylines
\setlength{\parindent}{0pt}
\textbf{Prompt Instruction:}
You are a helpful and precise assistant tasked with summarizing a web designer's decision steps in a think-aloud session. The designer conducted a think-aloud session while designing a web interface in Figma, explaining their thought process, decisions, and adjustments.

You will be provided with:
\begin{itemize}
  \item The full transcript (\texttt{Full transcript}) of the session (for overall session context), which includes the \texttt{decision\_step}'s transcript.
  \item A specific segment of the transcript containing the designer's decision steps (\texttt{Decision step}).
  \item Related task workflow screenshots (\texttt{Screenshots related to the decision step}) — \emph{note: these may or may not be provided for every decision step}.
  \item A clarification question generated to better understand the designer's rationale (\texttt{Question}).
  \item The designer's answer to that question (\texttt{User's Answer}).
\end{itemize}

Your Task
First, review the given \texttt{Decision step}, the full transcript, and the designer's answer to understand the workflow and clarify their decision rationale. Then, summarize the designer's key decision step into three clear, single-line sentences, each capturing the following:

\begin{enumerate}
  \item \textbf{Main Design Decisions and Actions} (\texttt{decision\_and\_actions} in the output format):  
  A single-line sentence summarizing the core design decisions and actions taken by the designer.

  \item \textbf{Rationale} (\texttt{rationale} in the output format):  
  A single-line sentence explaining the reason behind the designer's key decisions, incorporating insights from the user's answer.

  \item \textbf{Progression} (\texttt{progression} in the output format):  
  A single-line sentence describing how the design process logically progressed based on the designer's decisions.
\end{enumerate}

Guidelines
\begin{itemize}
  \item Focus on the core design decisions and actions taken by the designer with the explained rationale.
  \item Include key details essential to understanding what the designer did and why — such as specific properties adjusted (e.g., font size, spacing, layout), components added or removed, or designer's deep detailed reasoning process.
  \item Do not include extraneous details or unrelated information in the summary.
  \item Use clear and precise language to convey the main points of the designer's workflow.
  \item Ensure that the summary reflects the logical flow and sequence of the decision steps.
\end{itemize}

\textbf{Input:}
\begin{verbatim}
transcript: {transcript}
decision_step: {decision_step}
related_screenshots: {related_screenshots}
clarification_question: {clarification_question}
answer: {answer}
\end{verbatim}

\textbf{Formatting Instruction}:
\begin{verbatim}
{
    "decision_and_actions": "<decision_and_actions>",
    "rationale": "<rationale>",
    "progression": "<progression>"
}
\end{verbatim}
\end{prompt}

\section{Experiment Details}
\subsection{Next Step Generation Output in Section \ref{application:nextStep}}
\label{appendix:nextStep_example}
\subsubsection{Example 1.\\}
First step:
\begin{itemize}
    \item ``\textit{Decision and actions}'': The designer recognized the importance of temporal information and decided to add a date to the news section. They inserted the text "MAR, 2025" into the design. This text was placed in a text node with the ID "4006:191", which is part of a frame with the ID "4006:174". To accommodate the new content, the width of the text was adjusted from 0 to 250.
    \item ``\textit{Progression}'': By adding the date "MAR, 2025", the designer established a temporal context for the news section, which is essential for conveying timely information. This step lays the foundation for additional news-related content, logically advancing the design by allowing users to easily identify the time frame of the presented news.
\end{itemize}
Second step:
\begin{itemize}
    \item ``\textit{Decision and actions}'': The designer decided to change the font of the text elements to better align with the HCI research theme. Initially, the text elements "MAR, 2025" and "NEWS" used the "Passion One" font. The designer changed the font of both elements to the bold style of "Inter". The width of the "MAR, 2025" text was adjusted from 125 to 174, and the height from 35 to 39. The width of the "NEWS" text was changed from 132 to 195, and the height from 70 to 77. Later, the "NEWS" text style was further updated to "Inter Black", and its width was slightly adjusted from 195 to 198.
    \item ``\textit{Progression}'': The design process progressed by refining the typography to suit the clean and professional aesthetics suitable for HCI research. The designer’s choice of the "Inter" font—known for its clarity and modern appearance—helped improve the readability and visual appeal of the text elements. By adjusting the font style and size, the text was made more prominent and aligned with the overall design theme, contributing to a consistent and polished interface.
\end{itemize}
Generated step (Baseline):
\begin{itemize}
    \item ``\textit{Decision and actions}'': The designer decided to add a brief description below the "NEWS" heading to provide context for the news section. They inserted the text "Latest updates and insights from the lab" into a new text node with the ID "4006:200". This text was placed directly beneath the "NEWS" heading, with a top margin of 20 pixels to ensure visual separation. The font used was the regular style of "Inter", set at 16px, and the text color was a dark gray (\#333333) to maintain readability against a white background.
    \item ``\textit{Progression}'': This step enhances the content of the news section by offering users more information and encouraging engagement, thereby advancing the design. By providing context, the designer enables users to immediately understand the relevance of the news items, helping maintain user interest and facilitating website navigation. This decision also aligns with the overall goal of creating a consistent and user-friendly interface.
    \item ``\textit{Rationale}'': Adding a brief description below the "NEWS" heading improves clarity and enhances the user experience by helping users quickly understand the section's purpose. The choice of the regular "Inter" font ensures consistency with the existing typography, while the dark gray color provides a soft contrast that is easy on the eyes. The 20-pixel margin maintains a clean and organized layout by visually separating the description from the heading.
\end{itemize}

Generated step (condition without Answer)
\begin{itemize}
    \item ``\textit{Decision and actions}'': The designer decided to add a brief description below the "NEWS" heading to provide context for the news section. They inserted the text "Latest updates and insights from the lab" beneath the "NEWS" text element. The new text was styled using the "Inter" font in a regular weight, with a font size set to 18. To ensure proper alignment with the "NEWS" heading, the width of the text box was adjusted to 250.
    \item ``\textit{Progression}'': This step enriches the content of the news section by providing more information and making it more engaging, thereby advancing the design. By offering a brief description, the designer helps users anticipate the type of content they will find, improving site navigation and user engagement.
    \item ``\textit{Rationale}'': Adding a description beneath the "NEWS" heading enhances clarity by allowing users to quickly understand the section’s purpose, thereby improving the overall user experience. The choice of the "Inter" font maintains consistency with the existing typography, ensuring a unified appearance. The font size of 18 was selected to ensure readability while avoiding visual dominance over the primary "NEWS" heading.
\end{itemize}

Generated step (condition with Answer)
\begin{itemize}
    \item ``\textit{Decision and actions}'': The designer decided to provide context for the news items by adding a brief description below the "NEWS" section. They inserted the text "Improving student feedback from an HCI perspective" below the "MAR, 2025" date. This text was placed in a new text node with the ID "4006:200", and its width was set to 300 to fit well within the layout. The font used was "Inter Regular" with a size of 18, ensuring consistency with existing design elements.
    \item ``\textit{Progression}'': The design process progressed by enhancing the informational content of the "NEWS" section. After establishing a temporal context through the date, the designer logically added a descriptive element to provide greater depth and context. This step improves the overall user experience and engagement with the news section by helping users recognize not only the timing but also the focus of the content.
    \item ``\textit{Rationale}'': The reason for adding a description below the "NEWS" section was to give users a quick overview of the news content, enhancing understanding and engagement. By providing a concise summary, users can quickly grasp the relevance and focus of the news items—crucial for maintaining interest and ensuring accessible information. The choice of font and size aligns with the overall design theme, maintaining visual consistency and readability.
\end{itemize}

\subsubsection{Example 2.\\}
First step:
\begin{itemize}
    \item ``\textit{Decision and actions}'': The designer began by creating a new frame in Figma named "MacBook Pro 16 - 1". This action set up a workspace that mimics the screen size of a 16-inch MacBook Pro and serves as the foundation for the wireframe design of the Dental Monster website.
    \item ``\textit{Progression}'': By creating a frame tailored to the dimensions of a 16-inch MacBook Pro, the designer established a base layout for replicating the Dental Monster website. This step is crucial for ensuring the design is responsive and accurately represents how the site will appear on larger screens, laying the groundwork for detailed design work to follow.
\end{itemize}
Second step:
\begin{itemize}
    \item ``\textit{Decision and actions}'': The designer initially created a text element named "Text". They then changed the text content to "GENTLE MONSTER" and adjusted the width from 0 to 110. The font was initially set to "Inter Regular", then changed to "jsMath-cmmi10", and later updated again to "Bebas Neue Regular". As a result, the width decreased from 110 to 64. Finally, the font size was increased from 12 to 24, adjusting the width to 127.
    \item ``\textit{Progression}'': The designer aimed to create a text element resembling the original Gentle Monster logo. They began by creating a text element and setting its content to "GENTLE MONSTER". The designer experimented with various fonts—starting with "jsMath-cmmi10" to give a serif-like feel—but ultimately chose "Bebas Neue Regular" for a more fitting style. By increasing the font size, they enhanced visibility and presence, aligning the text element with the intended aesthetic of the original logo.
\end{itemize}
Generated step (Baseline; without captured decision steps):
\begin{itemize}
    \item ``\textit{Decision and actions}'': The designer will add a navigation bar at the top of the frame. This involves creating a horizontal rectangle spanning the width of the frame (1440px) with a height of 60px. The rectangle will be positioned at the top of the frame. The designer will then add text elements for navigation links such as "Home," "Shop," "About," and "Contact" using the "Bebas Neue Regular" font at a size of 18px. These text elements will be evenly distributed within the navigation bar.
    \item ``\textit{Progression}'': With the logo in place, the designer is now focusing on building the basic navigation structure. This step logically follows logo creation and begins to define essential layout and user interface elements for site navigation. The addition of the navigation bar sets the stage for developing the site's content and functionality.
    \item ``\textit{Rationale}'': Adding a navigation bar is a crucial step in establishing the foundational structure of an e-commerce site. It enhances usability and user experience by allowing easy access to different sections of the site. The choice of "Bebas Neue Regular" for the navigation links ensures visual consistency with the logo, maintaining a unified brand identity. The font size and spacing of the text elements are designed to ensure readability and a clean, modern appearance.
\end{itemize}

Generated step (condition without rationale)
\begin{itemize}
    \item ``\textit{Decision and actions}'': The designer decided to add a navigation bar at the top of the frame. This involved placing a new rectangular element with a width of 1440 pixels and a height of 60 pixels at the top of the frame (x: 0, y: 0). The rectangle was filled with a light gray color (\#F5F5F5) to visually distinguish it from the rest of the content. Inside this navigation bar, the designer added three text elements labeled "Home," "Shop," and "Contact", each set in "Bebas Neue Regular" with a font size of 18. These text elements were evenly spaced within the navigation bar, positioned at (x: 20, y: 20) for "Home", (x: 120, y: 20) for "Shop", and (x: 220, y: 20) for "Contact".
    \item ``\textit{Progression}'': By creating the navigation bar, the designer established a clear and organized layout for the top section of the webpage. This step builds upon the earlier decision to create a recognizable logo, further developing the site’s visual identity and improving user navigation. The designer is now ready to add more content and interactive elements to the page to ensure a comprehensive and user-friendly design.
    \item ``\textit{Rationale}'': Adding the navigation bar is a critical step in establishing the foundational structure of an e-commerce site. By using a light gray background, the designer ensures the navigation bar is visually distinct yet subtle, allowing users to focus on the content below. The choice of "Bebas Neue Regular" for the text maintains consistency with the existing "GENTLE MONSTER" logo, reinforcing a unified visual identity. The even spacing of the navigation items enhances usability by providing users with clear and accessible navigation options.
\end{itemize}

Generated step (condition with rationale)
\begin{itemize}
    \item ``\textit{Decision and actions}'': The designer decided to position the "GENTLE MONSTER" text element at the top center of the frame. This involved moving the text from its current position (x: 0, y: 0) to a new location within the frame at (x: 640, y: 50), aligning it horizontally at the center.
    \item ``\textit{Progression}'': The design process progressed by first establishing foundational elements such as the frame and text styling. By centering the "GENTLE MONSTER" text, the designer set a clear focal point for the page, guiding the placement of additional elements and content. This step ensures that the design aligns with the brand's aesthetic, maintaining a consistent and professional appearance.
    \item ``\textit{Rationale}'': The reason for centering the "GENTLE MONSTER" text is to create a balanced and visually appealing layout. Placing the logo at the top center of the page is a common design practice that helps establish brand identity and ensures the logo is immediately visible to users. This placement also aligns with general user expectations around brand positioning, thereby enhancing the overall user experience.
\end{itemize}

\subsection{Design Artifact Creation with Tool-Calling Agent in Section \ref{application:mcp}}
\subsubsection{Prompt: Style guideline extraction from the captured decision steps}
\label{appendix:prompt_style_extraction}
\mbox{}\par
\begin{prompt}
\obeylines
\setlength{\parindent}{0pt}
\textbf{Prompt Instruction:}
You are a design analyst tasked with extracting design style information from a web designer's decision steps. Focus specifically on visual design elements, styling choices, and aesthetic decisions.

\#\# Your Task
Analyze the designer's decision steps and extract information about their design style preferences, focusing on:

1. Layout Patterns: How they organize and arrange elements
2. Typography Choices: Font sizes, weights, spacing, hierarchy
3. Color Usage: Color choices, contrast, visual emphasis
4. Spacing \& Sizing: Margins, padding, element dimensions
5. Visual Hierarchy: How they create emphasis and organization
6. Component Styling: Buttons, frames, borders, effects

\#\# Guidelines
- Focus on concrete design decisions and visual choices
- Extract specific values (font sizes, colors, dimensions) when mentioned
- Identify recurring patterns in their design approach
- Note any unique or distinctive styling preferences
- Keep descriptions concise and practical

\#\# Output Format
Provide a clear, readable text summary organized by categories. Use bullet points and be specific about design choices and values.

Example format:
LAYOUT PATTERNS:
• Prefers left-aligned text layouts
• Uses consistent 24px margins
• Creates clear visual sections with spacing

TYPOGRAPHY:
• Default font size: 16px for body text
• Headers: 24px-36px range
• Line height: 1.5x for readability

COLOR CHOICES:
• Primary text: \#000000 (black)
• Background: \#FFFFFF (white)
• Accent color: \#007BFF (blue)

SPACING \& SIZING:
• Consistent 16px spacing between elements
• Button height: 40px standard
• Container max-width: 1200px

\textbf{Decision steps for designer:}
\begin{verbatim}
{{decision_steps_text}}
\end{verbatim}
\end{prompt}

\subsubsection{Prompt: tool-calling agent}
\label{appendix:tool-calling_agent_prompt}
\mbox{}\par
\begin{prompt}
\obeylines
\setlength{\parindent}{0pt}
\textbf{Prompt Instruction:}
You are a reasoning and action-taking agent within a co-creative UI design assistant.
  Your task is to produce an array of tool (function) calls necessary to fulfill the user's request.
  Interpret the user's intent and use the shared Figma canvas to explore, build, and iterate on design ideas.
  Include every necessary tool (function) calls and do not output any text other than the function calls themselves.

  1. Analyze the user's request
  - Understand what the user wants to achieve
  2. Be exhaustive
  - Consider all parameters, options, ordering, and dependencies necessary to fulfill the request in one turn.
  - You are allowed — and encouraged — to act proactively.

  Your task is to determine whether any design actions are needed based on:
  - the user's message:
    \${task\_instruction}
  - the current Figma canvas and its snapshot image
  - the results of previous tool operations
  \${feedback ? `
  - Previous Step Feedback: 
    \${feedback}
    
    This feedback was generated by analyzing the progress made in the previous operations. It compares what was accomplished versus what remains to be done according to the user's request. Use this feedback to:
    - Continue building on completed elements
    - Address the remaining tasks identified
    - Ensure you're progressing toward the user's request
    - Avoid repeating work that was already completed
  ` : ''}

   Please use the user's task instruction as the main criterion to determine the next action. 

  ---

  Tool Usage Rules
  - Use tools proactively and strategically to fulfill the user's request.
  - Leverage as many different tools as necessary to fully realize the user's requirements — not just minimal steps.
  - You may combine multiple tools in sequence if it helps achieve more complete or precise results.
  - If a previous tool call was incomplete, analyze the failure or gap, then take appropriate follow-up actions using additional tools.
  - If a previous tool call was complete, avoid unnecessary repetition unless the user's intent has changed.
  - If the current canvas already satisfies the user's intent, or if available tools are insufficient for meaningful action, you may choose not to act — but always verify thoroughly before skipping.

  Figma tool operation guidelines:  
      1. Figma Tool Basics
    - Each design element (node) is either a container (Frame/Component) or a leaf (Shape/Text).
    - Nodes expose uniform structural properties regardless of type.
    - Coordinates are global, relative to the canvas origin (0, 0) at the top-left corner.
    - Do not assume local coordinates unless explicitly provided.
    
    2. Node Hierarchy
    - All nodes belong to a single rooted tree mirroring the layer list.
    - Parent-child relationships define the hierarchy; a node's index in its parent determines both z-order and sidebar order.
    - If a child's size exceeds its parent container's bounds and the parent has 'clipsContent' enabled, the excess will not be visible.
    - If an element is placed inside a frame but positioned outside the frame's bounds (while 'clipsContent' is enabled), it will be unobservable.
    - Do not place nodes directly at the root unless they are intentional top-level elements.
    
    COMPONENT CREATION RULES:
    - For buttons: Create a frame with background color, corner radius, and auto-layout, then add text inside it. Do NOT create separate rectangle shapes inside the frame.
    - For logos: Create a frame with the logo's background shape (color, border radius, etc.), then add text or vector elements inside it. The frame itself should be the logo's visual container.
    - For cards: Create a frame with background and styling, then add content (text, images) inside it.
    - For navigation bars: Create a frame with background, then add navigation items inside it.
    - The frame itself should be the visual component, not just a container for separate shapes.
    
    POSITIONING EFFICIENCY:
    - Plan element positions before creating them to minimize move\_node calls.
    - Use Auto Layout strategically - if you enable it, let it handle positioning automatically.
    - Avoid excessive repositioning - create elements in their intended locations when possible.
    - Only use move\_node when absolutely necessary for final adjustments.
    
    3. Container Layout (Auto Layout)
    - Use frames strategically for components that need to contain other elements (e.g., buttons with text, cards with content, navigation bars).
    - IMPORTANT: When creating interactive components like buttons, the frame itself should serve as the visual shape/background, not as a separate container.
    - For buttons: Create a frame with the desired background color, corner radius, and styling, then add text directly inside it using auto-layout.
    - Avoid creating separate shape elements inside frames for simple components - let the frame be the component's visual representation.
    - When Auto Layout is applied, Figma automatically manages direction, gap, padding, and resizing of the container and its children.
    - Manual size or position edits must account for Auto Layout's automatic adjustments.
    - Enable Auto Layout only when certain about the intended layout behavior; otherwise, it may cause unexpected shifts.
    - When adding children to an Auto Layout container, verify that spacing and alignment remain correct.
    - CRITICAL: If content overflows outside the container and becomes hidden, immediately resize the frame so that all content remains visible. Never leave child nodes partially or completely hidden.
    
    AUTO LAYOUT POSITIONING RULES:
    - NEVER use 'move\_node' inside Auto Layout frames - Auto Layout will override absolute positioning.
    - To reorder elements in Auto Layout: Use 'move\_node\_into\_frame' with the 'index' parameter to specify the desired position (0 = first, 1 = second, etc.).
    - To change stacking order: Use 'reorder\_node' with 'TOP', 'BOTTOM', 'FORWARD', or 'BACKWARD' directions.
    - To adjust spacing: Modify the Auto Layout frame's gap, padding, or alignment properties instead of moving individual nodes.
    - If you need precise positioning: Consider disabling Auto Layout temporarily, make adjustments, then re-enable it.
    
    FRAME VISIBILITY RULES:
    - Ensure child nodes are visible when they should be: If nodes inside a frame are cut off or hidden when they should be visible, use 'resize\_node' to increase the frame's width and height to accommodate all content.
    - Check frame bounds when adding elements: When adding elements to a frame, verify that the frame's width and height are sufficient to display all children. Use 'resize\_node' if adjustments are needed.
    - Use 'clipsContent' wisely: If a frame has 'clipsContent' enabled, ensure the frame is large enough to show all intended content by using 'resize\_node'.
    - Adjust frame size when content is hidden: After adding new elements to a frame, use 'resize\_node' to expand the frame dimensions if any content becomes hidden or cut off.
    - Consider padding: Account for frame padding when calculating required frame dimensions. Use 'resize\_node' to add extra space for padding.
    - MANDATORY: If any child element is not fully visible within its parent frame when it should be visible, you MUST use 'resize\_node' to increase the frame's dimensions until all content is completely visible.
    
    4. Text Mechanics
    - Text nodes have font family, style, size, and other typographic properties independent of layout.
    - Resizing a text node does not scale the text; overflow text is clipped or wrapped based on settings.
    - Always check node size and alignment to avoid unintended overflow.
    - When placing text inside Auto Layout, ensure padding and alignment work together with text constraints.
    
    5. Naming Rules
    - Every node must have a unique, descriptive name.
    - Do not assign the same name to multiple nodes.
    - If a name already exists, generate a unique variant (e.g., 'btn\_primary\_v2').
    - Use consistent naming patterns for similar element types (e.g., 'btn\_primary', 'btn\_secondary', 'txt\_title').
    
    6. Preserve Existing Structure
    - Avoid unnecessary changes to nodes unrelated to the current task.
    - Do not delete, rename, or move unrelated layers.
    
    7. Property Changes Must Be Intentional
    - Only modify properties (size, color, position, etc.) relevant to the requested change.
    - Keep unrelated attributes (e.g., shadows, strokes) intact.
    
    8. Maintain Consistency
    - If multiple elements share a style (color, typography, size), maintain that consistency unless explicitly instructed otherwise.
    
    9. Avoid Overlapping or Hidden Elements
    - Ensure new elements do not unintentionally overlap or become hidden behind other layers.
    
    10. Validate Final Layout
    - After modifications, confirm that hierarchy, naming, alignment, and spacing remain correct and consistent.

  ---

  DESIGN STYLE GUIDELINES:
  Follow these specific design preferences to maintain consistency:

  \${\textbf{designStyle from captured decision steps}}

  ---
  \${feedback ? `
  Important - Feedback-Based Messaging:
  - Since previous step feedback exists, you MUST acknowledge what was completed and what remains incomplete
  - If there are remaining tasks in the feedback, explicitly mention that those aspects are not yet complete
  - Your message should reflect awareness of the incomplete elements identified in the feedback
  - Example: "Added the navigation frame as requested, but still need to complete the text labels and styling mentioned in the feedback."}
\end{prompt}
\subsubsection{Prompt: Feedback modul prompt for ReAct Agent}
\label{appendix:feedback_module_ReAct}
\mbox{}\par
\begin{prompt}
\obeylines
\setlength{\parindent}{0pt}
\textbf{Prompt Instruction:}
You are a feedback module for a design agent.
  
Your task is to analyze progress made toward the user's request by comparing the previous and current canvas contexts.

**User Request**: \${task\_instruction}

**Instructions**:
- Compare the previous canvas context with the current canvas context
- Analyze what progress was made toward achieving the user's request
- Identify what has been completed and what remains to be done
- Focus on user-request-relevant changes and progress
- **If minimal or no visual changes occurred**: Explicitly identify what specific actions need to be taken to make meaningful progress toward the user's request
- **PRIORITY: Text and Layout Issues**: Give HIGHEST priority to detecting and reporting text overlapping, text cut-off, and layout organization problems

**CRITICAL LAYOUT ISSUES TO DETECT**:
- **Frame Visibility Problems**: Check if any child elements inside frames are cut off, hidden, or not fully visible when they should be visible. If so, mention that frame resizing is needed.
- **Layout Organization Issues**: Look for elements that are poorly positioned, overlapping inappropriately, or not properly aligned within their containers.
- **Auto Layout Conflicts**: Identify if elements are being moved repeatedly or positioned incorrectly within Auto Layout frames.
- **Component Structure Problems**: Check if buttons, logos, or cards are created with unnecessary nested shapes instead of using frames as the visual component.
- **Text Overlap Issues**: Look for text elements that are overlapping with other elements or positioned too close together, making them unreadable.
- **Text Visibility in Components**: Check if text inside buttons, logos, or other components is fully visible and properly contained within the component boundaries.

**HIGH PRIORITY TEXT ISSUES**:
- **Text Overlapping**: If any text elements overlap with other text or UI elements, this is a CRITICAL issue that must be addressed immediately.
- **Text Cut-off in Buttons**: If button text is partially hidden or cut off, this makes the button unusable and must be fixed.
- **Text Cut-off in Logos**: If logo text is not fully visible, the brand identity is compromised.
- **Text Clipping in Frames**: If any text is clipped by frame boundaries when it should be fully visible, this affects readability.

**SPECIFIC FRAME RESIZING GUIDANCE**:
- **Horizontal Clipping**: If any component/element is cut off horizontally (left or right edges not visible), specify "Increase the frame's width using 'resize\_node' to show the full element".
- **Vertical Clipping**: If any component/element is cut off vertically (top or bottom edges not visible), specify "Increase the frame's height using 'resize\_node' to show the full element".
- **Both Directions**: If elements are clipped in both directions, specify "Increase both width and height of the frame using 'resize\_node' to accommodate all content".
- **Multiple Elements**: If multiple elements are clipped, specify which frame needs resizing and by how much (e.g., "Hero Section frame needs width increased by 200px").

**Output format**:
Return your analysis in exactly this format:
- **Completed**: [One clear sentence about what was accomplished]
- **Remaining**: [One clear sentence about what still needs to be done]

**IMPORTANT**: If you detect any of the critical layout issues mentioned above, include them in the "Remaining" section with specific recommendations. Examples:
- "Frame needs resizing to show all content" → "Hero Section frame needs width increased by 150px using 'resize\_node' to show the full button text"
- "Button should use frame as background instead of separate shapes" → "CTA button should be created as a single frame with background color instead of frame + rectangle + text structure"
- "Elements need proper Auto Layout positioning" → "Use 'move\_node\_into\_frame' with index parameter instead of 'move\_node' for Auto Layout frame positioning"
- "Text overlapping detected" → "CRITICAL: 'Get Started' button text is overlapping with subtitle text - separate these elements immediately"
- "Button text cut off" → "CRITICAL: CTA button text is partially hidden - increase button frame height by 20px using 'resize\_node'"
- "Logo text not visible" → "CRITICAL: Logo text is cut off at edges - increase logo frame dimensions using 'resize\_node' to show full text"

**Requirements**:
- Each item must be exactly ONE sentence
- Be specific and concise
- Focus on the most significant progress and next steps
- **MANDATORY**: If you detect ANY text overlapping, text cut-off, or layout organization issues, these MUST be mentioned in the "Remaining" section as CRITICAL priorities
- **Text issues take precedence**: Even if other progress was made, text visibility and layout problems must be addressed first

**NEVER ignore**: 
- Text overlapping, text cut-off, or layout organization issues - these are always critical You should feedback on these issues in the "Remaining" section.
\end{prompt}

\subsubsection{Tool List}
\label{appendix:tool_list}
\mbox{}\par
\begin{prompt}
\obeylines
\setlength{\parindent}{0pt}
1. Creation Tools (10 tools)  
create\_rectangle - Generate rectangular shapes with customizable size, color, stroke, and corner radius  
create\_frame - Create frame containers with Auto Layout support for organized layouts  
create\_frame\_from\_node - Convert existing nodes into frame containers  
create\_text - Generate text elements with font, size, alignment, and styling options  
create\_graphic - Create vector graphic elements for illustrations and icons  
create\_ellipse - Generate circular and elliptical shapes  
create\_polygon - Create polygonal shapes with configurable sides  
create\_star - Generate star-shaped elements with customizable points  
create\_line - Create linear elements and connectors  
create\_mask - Generate mask elements for visual effects  

2. Operation Tools (11 tools)  
move\_node - Relocate nodes to new positions with optional parent container changes  
move\_node\_into\_frame - Move nodes into specific frames with index positioning  
clone\_node - Duplicate nodes with optional new location or parent assignment  
resize\_node - Adjust node dimensions while maintaining aspect ratios  
delete\_node - Remove nodes from the canvas  
group\_nodes - Combine multiple nodes into organizational groups  
ungroup\_nodes - Break apart existing group containers  
rename\_node - Modify node display names for better organization  
rotate\_node - Apply rotation transformations around node centers  
boolean\_nodes - Perform vector boolean operations (union, subtract, intersect, exclude)  
reorder\_node - Modify z-order stacking (bring to front/back, layer adjustments)  

3. Style Tools (8 tools)  
set\_fill\_color - Configure solid background colors using RGBA values  
set\_corner\_radius - Adjust corner rounding for rounded rectangles  
set\_opacity - Control node transparency levels  
set\_stroke - Configure border styles including color, weight, and patterns  
set\_fill\_gradient - Apply gradient background fills with multiple color stops  
set\_drop\_shadow - Add drop shadow effects with customizable parameters  
set\_inner\_shadow - Apply inner shadow effects for depth  
set\_blend\_mode - Configure blend modes for visual composition  

4. Layout Tools (5 tools)  
set\_padding - Configure internal spacing for Auto Layout frames  
set\_axis\_align - Set primary and counter-axis alignment for Auto Layout  
set\_layout\_sizing - Configure horizontal and vertical sizing modes  
set\_item\_spacing - Control spacing between Auto Layout children  
set\_layout\_mode - Enable/disable Auto Layout with direction control  

5. Text Tools (4 tools)  
set\_text\_content - Modify text content of text nodes  
set\_text\_properties - Configure text attributes (size, color, alignment)  
set\_text\_decoration - Apply text decorations (underline, strikethrough)  
set\_text\_font - Configure font family, style, and weight properties  

6. Inspection Tools (7 tools)  
get\_page\_info - Retrieve current page information and top-level nodes  
get\_selection\_info - Get detailed information about selected nodes  
get\_node\_info - Query specific node properties and attributes  
get\_node\_info\_by\_types - Filter nodes by type and retrieve information  
get\_text\_node\_info - Extract text-specific node information  
get\_result\_image - Capture canvas as image for export  
get\_page\_structure - Analyze hierarchical page structure  

7. Component Tools (4 tools)  
create\_component - Convert existing nodes into reusable components  
create\_instance - Generate component instances with override capabilities  
create\_component\_set - Create component sets with variant management  
detach\_instance - Break component instance connections  

Total Tool Count: 49 tools  
Creation: 10 tools (shapes, text, frames, graphics)  
Operation: 11 tools (movement, resizing, cloning, deletion, grouping)  
Style: 8 tools (colors, effects, styling properties)  
Layout: 5 tools (Auto Layout control and configuration)  
Text: 4 tools (text manipulation and formatting)  
Inspection: 7 tools (information retrieval and analysis)  
Component: 4 tools (component system management)  

\end{prompt}

\end{document}